\documentclass{article}

\usepackage[english]{babel}

\usepackage[letterpaper,top=2cm,bottom=2cm,left=2.5cm,right=2.5cm,marginparwidth=1.75cm]{geometry}

\usepackage{amsmath}
\usepackage{amssymb}

\usepackage{graphicx}
\usepackage[colorlinks=true, allcolors=black]{hyperref}
\usepackage{bm}
\usepackage{float}
\usepackage{hyperref}
\usepackage{verbatim}
\usepackage[table]{xcolor}
\usepackage{subfigure}
\usepackage{caption}
\usepackage{multirow}
\usepackage{booktabs}
\usepackage{natbib}
\usepackage[flushleft]{threeparttable}
\usepackage{pdflscape}
\usepackage{everypage}
\usepackage{textcomp}
\usepackage{setspace}
\usepackage{color}

\title{\LARGE \bf Bayesian shared parameter joint models for heterogeneous populations \bigskip}

\author{\normalsize\textbf{Sida Chen}$^{*}$, \textbf{Danilo Alvares}, \textbf{Marco Palma}, \textbf{Jessica K. Barrett} \bigskip \\ \normalsize
MRC Biostatistics Unit, University of Cambridge, U.K. \\ \normalsize
$^{*}${\it email:} sida.chen@mrc-bsu.cam.ac.uk \\ \normalsize
}

\date{}

\begin{document}

\maketitle

\begin{abstract}
\noindent Joint models (JMs) for longitudinal and time-to-event data are an important class of biostatistical models in health and medical research. When the study population consists of heterogeneous subgroups, the standard JM may be inadequate and lead to misleading results. Joint latent class models (JLCMs) and their variants have been proposed to incorporate latent class structures into JMs. JLCMs are useful for identifying latent subgroup structures, obtaining a more nuanced understanding of the relationships between longitudinal outcomes, and improving prediction performance. We consider the generic form of JLCM, which poses significant computational challenges for both frequentist and Bayesian approaches due to the numerical intractability and multimodality of the associated model’s likelihood or posterior. Focusing on the less explored Bayesian paradigm, we propose a new Bayesian inference framework to tackle key limitations in the existing method. Our algorithm leverages state-of-the-art Markov chain Monte Carlo techniques and parallel computing for parameter estimation and model selection. Through a simulation study, we demonstrate the feasibility and superiority of our proposed method over the existing approach. Our simulations also generate important computational insights and practical guidance for implementing such complex models. We illustrate our method using data from the PAQUID prospective cohort study, where we jointly investigate the association between a repeatedly measured cognitive score and the risk of dementia and the latent class structure defined from the longitudinal outcomes.

\medskip

\end{abstract}

\section{Introduction} \label{sec:intro}

In many clinical and epidemiological studies, repeated measurements of biological markers and time-to-event data are simultaneously collected. For instance, in cohort studies on chronic conditions such as cardiovascular disease and lung disease, there has been great interest in characterizing the association patterns between marker trajectories and key health events, such as disease progression or death \citep{barrett2019estimating,su2021risk}. Joint models (JMs) have become an increasingly popular statistical modelling framework for jointly analyzing such longitudinal and time-to-event outcomes and demonstrate great potential for advancing medical insights and supporting patient monitoring \citep{rizopoulos2012joint,papageorgiou2019overview}.

In some scenarios, such as multi-center studies, the study population may naturally consist of heterogeneous sub-populations, with each one exhibiting potentially different longitudinal and time-to-event profiles. It may also be the case that there is additional heterogeneity that cannot be well explained by the recorded variables. In these situations, the standard JM may no longer be appropriate, and a naive implementation could lead to misleading inference results and conclusions. Joint latent class models (JLCMs) arise as an alternative framework to the standard JM, which explicitly model the latent subgroup structure by adopting a mixture model approach (see \citet{proust2014joint} for an overview). A fundamental assumption underlying the basic JLCM is the conditional independence of the longitudinal and time-to-event outcomes given the latent class membership, in other words, that implies that the association between the two outcome processes are fully explained via the shared latent class. While this assumption greatly alleviate the computational burden in model estimation, it no longer offers straightforward interpretation on the association structure between the longitudinal and time-to-event outcomes, which would be of interest in many clinical studies. Additionally, violation of this assumption could potentially necessitate the use of a larger number of latent classes, making them less interpretable. Therefore, the basic JLCM is mostly useful for prediction problems. 

To improve modelling flexibility and interpretability, the basic JLCM has been extended in more recent works in order to introduce additional dependency between the outcomes within each class. This is achieved by sharing random effects across the longitudinal and time-to-event submodels, and we refer to the resulting models as shared parameter JLCMs (see, e.g., \citet{liu2015joint, liu2020semi, andrinopoulou2020, wong2022semiparametric}). These models, however, introduce new challenges to statistical inference due to the presence of high-dimensional discrete and continuous latent variables. In frequentist methods, estimation is performed by maximizing the marginal likelihood function with all latent variables integrated out. For shared parameter JLCMs, this can be computationally prohibitive due to the need for numerical approximations to handle the integrals over random effects, and optimization can be challenging due to the presence of multiple local maxima. For Bayesian methods, inference is based on the posterior distribution of model parameters, along with the latent variables, and numerical integration with respect to random effects is not required. Available information on the data can be incorporated via priors, which can also have a regularizing effect on inference. In this paper, we focus on the Bayesian paradigm due to its appeal from both modelling and computational perspectives. To our knowledge, the only existing Bayesian method is that proposed by \citet{andrinopoulou2020}, who introduced a two-stage estimation framework. In stage 1, an overfitted mixture (a mixture with potentially more classes than necessary) is employed to infer the number of latent classes by removing `empty' classes. In stage 2, conditional on the results from stage 1, a shared parameter JLCM is refitted. However, there are several limitations to this approach. One challenge with the overfitted mixture model is that it can be computationally difficult to fit due to its complexity. Additionally, the number of latent classes inferred is highly sensitive to a threshold parameter, which requires tuning. \citet{andrinopoulou2020} used classical Markov chain Monte Carlo (MCMC) methods, implemented in JAGS, for posterior sampling. However, with high-dimensional and multimodal posterior spaces, such as in this model, these MCMC methods may suffer from poor mixing and convergence issues. Another important issue not properly addressed in their work is the presence of multimodality in the posterior, even with label ordering constraints. This issue can be particularly significant when using standard noninformative priors, as is commonly done in JMs. In such scenarios, relying on a single MCMC chain can lead to biased inference. 

In this paper, we propose a new Bayesian inference framework for shared parameter JLCMs, addressing some of the key limitations of existing Bayesian approaches. Our method is built on a finite mixture model framework. We consider models with plausible numbers of latent classes. A more appropriate model for the given data can then be selected through Bayesian model comparison, which may be conducted using suitable information criteria. From the modelling perspective, we found that using mildly informative priors can be highly advantageous for shared parameter JLCMs, as they help with model identifiability and improve MCMC convergence. These priors can be informed by available domain knowledge or through an empirical Bayes approach. To perform posterior inference for a given model, differently from \citet{andrinopoulou2020}, we utilize the No-U-Turn Sampler (NUTS), an auto-tuned Hamiltonian Monte Carlo algorithm, for sampling all continuous parameters \citep{hoffman2014no}. NUTS is particularly advantageous when the posterior is of high dimensionality and complexity, and is widely regarded as the state-of-the-art MCMC method \citep{vstrumbelj2024past}. To appropriately handle the multimodal posterior, we propose a computationally simple yet effective approach that involves embarrassingly parallel MCMC sampling, leveraging multi-core computing. The effectiveness of our proposed method, and its superiority over the existing approach, is demonstrated through a simulation study that follows and extends the settings considered in \citet{andrinopoulou2020}. We further illustrate the method using data from the well-known PAQUID prospective cohort study \citep{letenneur1994incidence}, accessed via the {\tt lcmm} R package \citep{proust2023package}. In this analysis, we investigate the latent subgroup structure within the population and examine the relationship between cognitive aging and the risk of dementia.

The rest of this paper is structured as follows. In Section \ref{sec:model}, we present the formulation of the shared parameter JLCM. Section \ref{sec:inference} discusses several key aspects related to the Bayesian estimation of the joint model. In Section \ref{sec:sim}, we present a simulation study to compare our method with the existing approach and investigate the impact of a specific model misspecification. An application to data from the PAQUID cohort study is presented in Section \ref{sec:appl}. Finally, Section \ref{sec:disc} concludes with a discussion of the proposed method and potential future perspectives.

\section{Shared parameter joint latent class models} \label{sec:model}

In this section we present the model in its general form. We assume the population of sample size $n$ consists of a fixed $G$ latent classes, where $G$ may be unknown. Each latent class is characterized by a class-specific joint distribution of the longitudinal and time-to-event outcomes formed via shared random effects. The model consists of the following three submodel components.

\subsection{The class membership submodel}

Let $c_{i}$ denote the latent class membership indicator for individual $i$, $i=1,\ldots,n$, with $c_{i}=g$, $g=1,\ldots, G$, if subject $i$ belongs to the $g$th class. The indicator $c_{i}$ is unobserved and is modelled using a categorical distribution with a probability vector $\pi_{i} = (\pi_{i1},\ldots,\pi_{iG})$, where $\pi_{ig}=P(c_{i}=g)$. The membership probabilities are typically related to a set of $p$-dimensional time-independent baseline covariates $W_{i}$, via the softmax link function
\begin{equation}\label{eq:model_classmember}
    \pi_{ig} = P(c_{i}=g \mid W_{i}) = \frac{\exp(\psi_{g0}+W_{i}^{T}\psi_{g})}{\sum_{k=1}^{G}\exp(\psi_{k0}+W_{i}^{T}\psi_{k})}, \quad g=1,\ldots,G,
\end{equation}
where $\psi_{g}=(\psi_{g1},\ldots,\psi_{gp})$ are the class-specific vectors of regression coefficients associated with $W_{i}$. For identifiability purpose, we set $\psi_{G0}$ and $\psi_{G}$ to zeros. When no external covariates are considered, the membership probability vector becomes the same across all subjects, i.e., $\pi_{i}=\pi$ for all $i$. In this scenario, \citet{andrinopoulou2020} proposed an alternative model, where $\pi$ is directly modelled on the simplex using a Dirichlet distribution. However, with this approach, incorporating covariates is difficult, and as we will show in Section \ref{sec:sim}, the omission of relevant covariates can be problematic as it can introduce estimation bias to other model parameters.

\subsection{The longitudinal submodel}

Let $y_{i}(t)$ denote the value of the longitudinal outcome of subject $i$ measured at time $t$, and let $y_{i}=(y_{i1},\ldots,y_{in_{i}})$ be the observed $n_{i}$-dimensional longitudinal response vector, where $y_{ij}=y_{i}(t_{ij})$, $j=1,\ldots,n_{i}$. Conditional on the latent class $g$, the longitudinal outcome is typically modelled using a generalized linear mixed model (GLMM) framework. An important difference from the standard JM ($G=1$) is that we now need to introduce class-specific random effects $b_{ig}$ to characterize the deviation of a subject's marker trajectory relative to the average marker trajectory of class $g$. Assuming the same set of random effects across classes (i.e., $b_{i1}=b_{i2}=\ldots=b_{iG}$) can, therefore, be problematic and may lead to biased inference. Denote $b_{i}=(b_{i1},b_{i2},\ldots,b_{iG})$. Assuming that the $r$-dimensional random effect vector $b_{ig}$ is distributed according to a multivariate normal distribution $N(0,\Sigma_{g})$, then given $c_{i}=g$ and $b_{i}$, 
the response $y_{ij}$ are assumed to be independent and belong to a member of the exponential family with density
\begin{equation}\label{eq:model_ef}
    f(y_{ij} \mid c_{i}=g, b_i) = \exp\left\{\frac{y_{ij} \zeta_{ij}(b_{ig}) - a_{1}(\zeta_{ij}(b_{ig}))}{a_{2}(\eta_{g})} - a_{3}(y_{ij}; \eta_{g})\right\},
\end{equation}
where $\zeta_{ij}(b_{ig})$ and $\eta_{g}$ denote the natural and dispersion parameters in the exponential family, respectively, and $a_1(\cdot)$, $a_2(\cdot)$, and $a_3(\cdot)$ are known functions specifying the member of the exponential family. The conditional mean of $y_{ij}$ given the class membership and random effects is related to the linear predictors via
\begin{equation}\label{eq:model_long_mean}
    E[y_{ij}\mid c_{i}=g, b_{i}, X_i(t), Z_i(t)]=a_{1}'(\zeta_{ij}) =g^{-1}(X^{T}_{i}(t_{ij})\beta_{g}+Z^{T}_{i}(t_{ij})b_{ig}),
\end{equation}
where $a_{1}'(\cdot)$ denotes the first derivative with respect to its argument, $g(\cdot)$ denotes a known monotonic link function specified according to the member of the exponential family, $\beta_{g}$ is a vector of class-specific regression coefficients, and $X^{T}_{i}(t)$ and $Z^{T}_{i}(t)$ denote the possibly time-dependent design vectors for the fixed and random effects, respectively. In the case of continuous longitudinal response, the GLMM reduces to a standard Gaussian linear mixed model (LMM):
\begin{equation}
    (y_{ij} \mid c_{i}=g, b_{i}) \sim N(\mu_{i}(t_{ij} \mid c_{i}=g, b_{i}, X_i(t), Z_i(t)), \sigma^{2}_{g}),
\end{equation}
where $\mu_{i}(t_{ij} \mid c_{i}=g, b_{i}, X_i(t), Z_i(t))=X^{T}_{i}(t_{ij})\beta_{g}+Z^{T}_{i}(t_{ij})b_{ig}$ and $ \sigma^{2}_{g}$ is the class-specific error variance.

\subsection{The time-to-event submodel}

Let $T^{*}_{i}$ be the true event time of interest and $C_{i}$ be the censoring time for subject $i$. We define the observed event time as $T_{i}=\min(T^{*}_{i}, C_{i})$, and the censoring indicator variable $\Delta_{i}=I(T^{*}_{i}\leq C_{i})$, where $I(\cdot)$ is the indicator function.  Conditional on the latent class membership, we model the risk of the event via the class-specific hazard function as
\begin{equation}\label{eq:model_haz}
    h_{i}(t \mid c_{i}=g, b_i, \Tilde{W}_{i})=h_{0g}(t)\exp\left\{\Tilde{W}_{i}^{T}\gamma_{g}+f(\beta_{g},b_{ig},t,\alpha_{g})\right\},
\end{equation}
where $h_{0g}(t)$ is the baseline hazard function parameterized by $\phi_g$, $\gamma_{g}$ denotes the vector of regression coefficients associated with the vector of exogenous risk factors $\Tilde{W}_{i}$, and $f(\cdot)$ serves the role to link the longitudinal and the time-to-event processes within a latent class. In this paper, we consider the so-called current value association structure, where $f(\cdot)=\alpha_{g}\mu_{i}(t_{ij} \mid c_{i}=g, b_{i}, X_i, Z_i)$, with $\alpha_{g}$ characterizing the strength of the association. This is perhaps the most popular option in the JM literature but other reasonable functional forms may be considered, depending on the application context. Note that without the inclusion of the function $f(\cdot)$, \eqref{eq:model_haz} reduces to a standard proportional hazard model and the resulting full model becomes the basic JLCM of \citet{proust2014joint}.

\section{Bayesian inference}\label{sec:inference}

\subsection{The Bayesian model}\label{subsec:model}

Let the full dataset $D^{(n)}=(D_{1},\ldots,D_{n})$, where $D_{i}=(y_{i},T_{i},\Delta_{i})$, $c=(c_{1},\ldots,c_{n})$, $\bm{b}=(b_{1},\ldots,b_{n})$, and let $\Theta=(\{\psi_{g0}\}_{g=1}^{G-1},\{\psi_g\}_{g=1}^{G-1}, \{\eta_g\}_{g=1}^{G}, \{\beta_g\}_{g=1}^{G}, \{\phi_g\}_{g=1}^{G}, \{\alpha_g\}_{g=1}^{G}, \{\gamma_g\}_{g=1}^{G}, \{\Sigma_g\}_{g=1}^{G})$ denote the collection of all parameters associated with the model presented in Section \ref{sec:model}. We assume that the joint posterior density of model parameters, random effects and latent class indicator variables, takes the form
\begin{align}\label{eq:posterior}
\begin{split}
    p(\Theta, b, c \mid D^{(n)}) &\propto\prod_{i=1}^{n} \left(p(D_i \mid b_i, c_i, \Theta)p(c_i \mid \Theta)\prod_{g=1}^{G}p(b_{ig} \mid \Theta)\right) \times p(\Theta)\\
    &= \prod_{i=1}^{n} \left( \prod_{g=1}^{G} \left( p(y_i \mid b_i, c_i = g, \Theta) p(T_i, \Delta_i \mid b_i, c_i = g, \Theta) \pi_{ig} \right)^{I(c_i=g)} p(b_{ig} \mid \Theta) \right) \times p(\Theta),
\end{split}
\end{align}
where $p(y_i \mid b_i, c_i = g, \Theta)$ is given by \eqref{eq:model_ef} and \eqref{eq:model_long_mean}, $\pi_{ig}$ is given by \eqref{eq:model_classmember}, and $p(b_{ig} \mid \Theta)$ is the density of multivariate normal distribution with mean 0 and covariance matrix $\Sigma_g$. $p(T_i, \Delta_i \mid b_i, c_i = g, \Theta)$ is the likelihood contribution from the survival submodel, which takes the form
\begin{equation}\label{eq:lik_surv}
    p(T_i, \Delta_i \mid b_i, c_i = g, \Theta) = h_{i}(t \mid c_{i}=g, b_i)^{I(\Delta_i=1)}\times \exp\left\{-\int_{0}^{T_i}h_{i}(s \mid c_{i}=g, b_i)ds\right\},
\end{equation}
where the hazard function $h_i$ is given by \eqref{eq:model_haz}. The integral in \eqref{eq:lik_surv} cannot be computed analytically in general and numerical approximation is required. In our implementation, we used the Gaussian-Legendre quadrature method (with 15 quadrature points).

We assume that the joint prior density $p(\Theta)$ is factorized into a product of prior density for each parameter. For the model in consideration, prior specification requires some care. Either overly strong or vague priors could be problematic: the former prior setting could have strong and undesirable influence on inference, such as on the selection of number of latent classes (see Section \ref{subsec:modelchoice}), especially when the sample size is small, whereas the latter setting may lead to a highly multi-modal posterior and MCMC could show mixing issues. To help with model identifiability and MCMC sampling efficiency, we propose to use mildly informative priors where possible. Hyperparameters can be selected based on available information for the data and expected structure for the resulting model, and/or guided by an empirical Bayes approach. 

We want to give special attention to the priors for the variance parameters given their important contribution to the posterior and the specific structure of the mixture model. Standard default noninformative choices, such as the inverse gamma prior with extremely small shape and scale parameters, may not be appropriate when $G>1$. When the LMM is used for the longitudinal submodel, we can use the popular half-Normal distribution as a prior on the residual error variance $\sigma_{g}^{2}$. Since $\sigma_{g}^{2}$ is not expected to be influenced much by the time-to-event data or varying values of $G$, the hyperparameter can be selected based on a LMM fit (with $G=1$) to the longitudinal data alone. Light-tailedness of the normal distribution is desirable to prevent excessively large values. In this paper, we assume the covariance matrices $\Sigma_g$ to be diagonal, so setting the prior for $\Sigma_g$ amounts to setting the prior for the random effect variances. Note that the random effects are still allowed to be correlated {\it a posteriori} if supported by the data, as there is no independence constraint in the posterior distribution of the random effects. Heuristically, we want the prior to prevent both excessively small and large values. Extremely small random effect variances may be associated with the formation of `small' clusters or cluster degeneracy, and mixing issues can occur when the variance approaches zero due to instability in the posterior. In contrast, large values could conflict with the latent classes' role in explaining population heterogeneity, coincide with the formation of a `large' class, and/or the collapse of other classes. In our implementation, we have found that a Gamma distribution with equal shape and rate parameters greater than 1 provides a good choice to cover variance values within a reasonable range. To get an idea of the plausible range of random effects variances for a specific model, we can fit a basic JLCM conditional on different candidate values of $G$ using the MLE approach developed in \citet{2017lcmm}, implemented via the {\tt lcmm} R package, noting that random effects variances are generally expected to decrease as $G$ increases. Later, in Sections \ref{sec:sim} and \ref{sec:appl}, we provide concrete examples and explanations regarding the choice of priors. 
 
\subsection{Posterior sampling}\label{subsec:mcmc}

The resulting posterior distribution defined via equation \eqref{eq:posterior} is not analytically tractable so we resort to MCMC methods for posterior inference. We consider the NUTS algorithm due to its efficiency for sampling from high dimensional parameter spaces. NUTS requires the underlying density to be smooth because it relies on gradient computation, so it cannot be used to directly sample from \eqref{eq:posterior} due to the presence of discrete variables. Instead, we use NUTS to sample from the posterior with $c_i$ marginalized out from \eqref{eq:posterior}, $p(\Theta, b \mid D^{(n)})$, which is proportional to
\begin{equation}
   \prod_{i=1}^{n} \left[ \left(\sum_{g=1}^{G}  p(y_i \mid b_i, c_i = g, \Theta) p(T_i, \Delta_i \mid b_i, c_i = g, \Theta) \pi_{ig} \right) \prod_{g=1}^{G}p(b_{ig} \mid \Theta) \right] \times p(\Theta).
\end{equation}

Posterior samples of latent class allocation variable $c_i$ can be obtained by sampling exactly from its full conditional distribution $p(c_i\mid \Theta^{(j)}, b^{(j)}, D^{(n)})$, which is a categorical distribution with weights for class $g$, $g=1,\ldots,G$, proportional to
\begin{equation}
    p(y_i \mid b_i^{(j)}, c_i = g, \Theta^{(j)}) p(T_i, \Delta_i \mid b_i^{(j)}, c_i = g, \Theta^{(j)}) \pi_{ig}^{(j)},
\end{equation}
where $b_i^{(j)}$ and $\Theta^{(j)}$ are the $j$th MCMC samples from $p(\Theta, b \mid D^{(n)})$, and $\pi_{ig}^{(j)}$ is computed using $\Theta^{(j)}$. Note that based on the sampled $c_i$, the posterior class membership probability can be estimated as 
\begin{equation}
    \hat{p}(c_i=g \mid D^{(n)})=\frac{1}{T}\sum_{j=1}^{T}I(c_{i}^{(j)}=g), \quad i=1,\ldots,n,\quad g=1,\ldots,G.
\end{equation}

\subsection{Tackling multimodality in the posterior}\label{subsec:multimodality}

For mixture-type models, an obvious cause of multimodality in the posterior is the invariance of the likelihood under permutations of the latent class labels. This can create problems for MCMC-based inference because label switching may occur (in theory) during MCMC runs, rendering the marginal posterior for class-specific parameters unidentifiable. To ensure valid inference, some kind of post-processing of MCMC samples is required, see, e.g., \citet{stephens2000dealing}. However, in complex hierarchical models like the one considered here, label switching is nearly impossible within a practical number of MCMC runs because it requires a simultaneous jump for all class-specific parameters. In our experiments, label switching did not occur for our models.

What needs to be tackled is that multiple well-separated local regions of high density can still appear in the posterior, even within the space of a specific class labelling. This is observed in MCMC runs (especially with the use of noninformative priors) and is overlooked in \citet{andrinopoulou2020}. This makes single-chain MCMC inference problematic, as the chain may get trapped in a local region depending on the location of initialization. Various MCMC schemes have been proposed in the literature to handle complex multimodal distributions, notably parallel tempering, which involves running multiple chains that communicate with each other. However, effective use of such algorithms requires careful design and tuning, and they can be computationally prohibitive for complex models. Our idea here is motivated by \citep{yao2022stacking}, who propose a Bayesian stacking approach to take advantage of fully parallel MCMC sampling, aiming to combine locally stuck chains to enhance prediction performance. When the focus is on inference, a modification leads to a sampling scheme with the following three steps:
\begin{enumerate}
    \item Run $M$ parallel chains with different initializations.
    \item  Clustering the $M$ parallel chains into $K$ clusters, defined over regions $\Omega_1, \ldots, \Omega_K$, with $K \leq M$.
    \item Select a region $\Omega_k$ with probability proportional to 
    \[
    w_k = \int P(D \mid b, \boldsymbol{\Theta}) P(b \mid \boldsymbol{\Theta}) P(\boldsymbol{\Theta}) \mathbb{I} \left( (b, \boldsymbol{\Theta}) \in \Omega_k \right) dbd\boldsymbol{\Theta},
    \]
    and then draw a sample with replacement from $\Omega_k$.
\end{enumerate}
In step 2, a between-chain mixing measure such as $\hat{R}$ can be used for clustering the chains \citep{vehtari2021rank}, and the weights $w_k$ can be estimated by any valid Monte Carlo estimator for the marginal likelihood based on samples in $\Omega_k$ \citep{llorente2023marginal}. Assuming that the chains cover all important regions of the support, the final samples would form approximate samples from the target posterior, as this is essentially a importance resampling procedure \citep{smith1992bayesian}. 

When the dimensionality of the parameter space becomes high, the clustering procedure in step 2 can still be challenging to implement. To improve computational feasibility, in our implementation, we consider a simplified version of the scheme above with the clustering step removed, and we select one among the $M$ parallel chains that produces the highest weights, upon which inference will be based.
This approach can be justified with the assumption that the posterior is dominated by a single region $\Omega_k^{*}$ in terms of posterior mass, which may be reasonable with the use of mildly informative priors, or, if we are willing to use the model that provides the largest probability given the data. In our simulations, we found this simplified scheme performed satisfactorily.
Note that the posterior within $\Omega_k$ may still have a complex structure; therefore, it remains important that the MCMC algorithm is efficient so that it adequately explores the local region, which motivates our choice of the NUTS algorithm. 
In our implementation, to estimate the weights $w_k$, we consider the truncated harmonic mean estimator \citep{Robert2009, chen2023bayesian}, which is given by 
\begin{equation}
    \left( \frac{1}{T} \sum_{i=1}^{T} \frac{h\left(b^{(i)}, \boldsymbol{\Theta}^{(i)}\right)}{p\left(D^{(n)} \mid b^{(i)}, \boldsymbol{\Theta}^{(i)}\right) p\left(b^{(i)}, \boldsymbol{\Theta}^{(i)}\right)} \right)^{-1},
\end{equation}
where $b^{(i)}$ and $\boldsymbol{\Theta}^{(i)}$ are the $i$th MCMC samples out of a total of $T$ samples. A convenient choice for $h$ is 
\begin{equation}
    h(b,\boldsymbol{\Theta}) = \frac{1}{V(\epsilon) \beta T} \sum_{j: (b^{(j)},\boldsymbol{\Theta}^{(j)}) \in \mathcal{H}_{\beta}} \mathbb{I}(d((b^{(j)},\boldsymbol{\Theta}^{(j)}), (b, \boldsymbol{\Theta})) < \epsilon),
\end{equation}
where $V(\epsilon)$ is the volume of the ball centered at $(b,\boldsymbol{\Theta})$ with radius $\epsilon$ (small), $\mathcal{H}_{\beta}=\{(b^{(j)},\boldsymbol{\Theta}^{(j)}): \\p(D^{(n)} \mid b^{(i)}, \boldsymbol{\Theta}^{(i)}) p(b^{(i)}, \boldsymbol{\Theta}^{(i)})>q_\beta\}$, and $q_\beta$ is the empirical upper $\beta$ quantile of $p(D^{(n)} \mid b^{(i)}, \boldsymbol{\Theta}^{(i)}) p(b^{(i)}, \boldsymbol{\Theta}^{(i)})$. When the radius is sufficiently small, $h(b^{(i)}, \boldsymbol{\Theta}^{(i)})$ will take the value of $1/(V(\epsilon) \beta T)$ if the sample is from $\mathcal{H}_{\beta}$, and zero otherwise. Note also that $V(\epsilon)$ does not need to be computed, as it will cancel out during normalization.
To further aid in convergence, we initialize the chains making use of the MLE obtained under the basic JLCM via the {\tt lcmm} R package. More specifically, initial values for $\eta_g, \beta_g, \phi_g, \gamma_g$ are motivated by the MLE, followed by a random perturbation. All other parameters and latent variables are randomly initialized.

\subsection{Selection of number of classes}\label{subsec:modelchoice}

In many application contexts, the number of latent classes $G$, if any, is unknown. To infer this value, we choose not to adopt the overfitted mixture approach proposed by \citet{andrinopoulou2020} due to the aforementioned difficulties. Instead, we propose simultaneously considering models with candidate values of $G$, treating it as a model comparison problem. One popular class of methods for Bayesian model choice is based on marginal likelihood or Bayes factors, which generally have consistency in model selection when the true generating process is among the candidate models, though it is worth noting that, in practice, seeking the `true' number of classes may not be meaningful or realistic. An important issue with these approaches is that the results can be sensitive to the choice of priors. For our model, we find that the prior for the random effect variances can have a strong influence on the estimated marginal likelihood.

Here, we consider a different class of methods to guide our choice on $G$, which are based on expected predictive error or generalization loss. The two most popular criteria for this purpose are the Bayesian leave-one-out information criterion (LOOIC) and the widely applicable information criterion (WAIC) \citep{vehtari2017practical}. LOOIC targets $\sum_{i=1}^{n}\log p(D_i \mid D_{-i})$ (or multiplied by $-2$ to be on the deviance scale), where in our context $p(D_i \mid D_{-i})=\int p(D_i \mid \bm{b}, \bm{c}, \Theta)p(  \bm{b}, \bm{c}, \Theta \mid D_{-i})d\bm{b}d\bm{c}d\Theta$. WAIC, on the other hand, targets $\sum_{i=1}^{n}\log p(D_i \mid D^{(n)})-p_{\text{WAIC}}$ (or multiplied by $-2$ to be on the deviance scale), where $p(D_i \mid D^{(n)})=\int p(D_i \mid \bm{b}, \bm{c}, \Theta)p(  \bm{b}, \bm{c}, \Theta \mid D^{(n)})d\bm{b}d\bm{c}d\Theta$, and $p_{\text{WAIC}}$ is defined as the sum of the posterior variance of the log predictive density for each data point.  To compare two models, we examine the significance of the difference in predictive performance by comparing the z-score, which is computed by dividing the estimated difference in LOOIC or WAIC by the estimated standard error associated with the difference, to the critical value for a one-tailed test based on the standard normal distribution at a given significance level \citep{sivula2020uncertainty}. 
The two criteria and the quantities required for computing the z-score can be estimated based on MCMC outputs using the method developed in \citet{vehtari2017practical}, which is implemented in the {\tt loo} R package \citep{loo}.
In line with the parsimony principle, a more complex model should be selected only if it demonstrates a statistically significant improvement. In our experience (see also results in Section \ref{sec:sim}), we found that a relatively stringent significance level, such as $\alpha=0.1\%$, is necessary to guard against the risk of overfitting. We want to emphasize that in practice, one should not blindly follow results based on such criteria, as they can be influenced by deviations in model formulation from the unknown true data-generating process (if one exists) and the quality of the estimates of these criteria. Therefore, these results should be considered with cautiousness in conjunction with model interpretation and the application context.

\section{Simulation study} \label{sec:sim}

We consider a simulation study consisting of two simulation settings. Setting I follows exactly from the simulation study considered in \citet{andrinopoulou2020}, with the aim of providing a `sanity check' for the proposed method and evaluating the performance of our approach in comparison to theirs. Setting II, modified from Setting I, aims to evaluate the impact of the misspecification of the class membership submodel on inference.

\subsection{Simulation settings}
\subsubsection{Setting I}
In this setting, three scenarios are considered for the latent class structure, with G equals one, two and three. The total sample size is fixed at $n=900$, and the latent class membership is randomly assigned to each individual according to pre-specified proportions of each class. The class-specific longitudinal trajectory is generated from a LMM specified as
\begin{equation}
    y_i(t) = \beta_{g0} + \beta_{g1} t + b_{ig0} + b_{ig1} t + \beta_{g2} \text{male}_i + \epsilon_{ig}(t),
\end{equation}
where $\epsilon_{ig}(t)\sim N(0, \sigma^{2}_{g})$, $(b_{ig0}, b_{ig1})\sim N(0,\Sigma_g)$, and $\text{male}_i$ is an artificial covariate for gender, simulated from a Bernoulli distribution with a probability 0.5 for being male (labelled as 1). The class-specific time-to-event model is specified as
\begin{equation}
    h_i(t) = \phi_{g0} t^{\phi_{g0} - 1} \exp\left\{ \phi_{g1} + \gamma_{g1} \text{age}_i + \alpha_g \mu_i(t) \right\},
\end{equation}
where $\mu_i(t)=\beta_{g0} + \beta_{g1} t + b_{ig0} + b_{ig1} t + \beta_{g2} \text{male}_i$. $\text{age}_i$ is generated from a normal distribution centered at 45 with a standard deviation of 15.70. For each individual, a random censoring time $C_i$ is generated from a uniform distribution between zero and 17.5. We refer to \citet{andrinopoulou2020} (Table 1) for further details on parameter value settings. For each scenario, we evaluated the performance of LOOIC and WAIC in selecting the number of classes by considering candidate values of $G$ ranging from one to four, and we considered three significance levels ($\alpha=5\%$, $1\%$, $0.1\%$) for performing the model comparison as described in Section \ref{subsec:modelchoice}. Conditional on the true value of $G$, we further examine the performance of the proposed algorithm in terms of accuracy in parameter estimation and classification. Under Setting I, the estimation model (for a given $G$) assumes the same model formulation for both the longitudinal and survival submodels as used in the generating process, with no covariates included in the membership submodel.

\subsubsection{Setting II}

 For this setting we have fixed $G=2$. The data generation process for the longitudinal and survival submodel are the same as in Setting I. We consider 4 different scenarios to generate the class membership:
\begin{itemize}
    \item Scenario 1: class allocation independent of any covariates (same as Setting I with $G=2$). 
    \item Scenario 2: $W_i=\Tilde{x}_i$, where $\Tilde{x}_i\sim N(0,1)$,  with $\psi_{10}=-0.4$, $\psi_{11}=1$.
    \item Scenario 3: $W_i=(\text{male}_i, \text{age}_i)$,  with $\psi_{10}=2$, $\psi_{11}=4$, $\psi_{12}=-0.1$.
    \item Scenario 4: $W_i=(\text{male}_i, \text{age}_i, \Tilde{x}_i)$,  with $\psi_{10}=2$, $\psi_{11}=4$, $\psi_{12}=-0.1$, $\psi_{13}=1$.
\end{itemize}

For each scenario, we consider two estimation models, which represent two common choices for modelling the class membership. Model 1 assumes homogeneous class membership probabilities as used in Setting I (i.e., correctly specified for Scenario 1), whereas model 2 considers the same set of covariates as considered in class-specific JM with $W_i=(\text{male}_i, \text{age}_i)$ (i.e., correctly specified for Scenario 3). Results from the two estimation models are compared in terms of the accuracy of estimating the commonly shared parameters, as well as classification accuracy.

\subsection{Estimation settings}

To implement the proposed algorithm, we use the following set of priors for the model parameters throughout: 
\begin{equation}\label{eq:priorsim}
\begin{gathered}
   \beta_{gi}, \gamma_{gi}, \phi_{g1},  \alpha_g \sim N(0,5^2), \quad \psi_{gi}\sim N(0,2^2), \quad \phi_{g0} \sim \text{Gamma}(2,0.5),\\
   \sigma^{2}_{g}\sim \text{half-Normal}(0,0.5^2), \quad \Sigma_{g,ii}\sim \text{Gamma}(1.5,1.5), \\
\end{gathered}
\end{equation}
where $\text{Gamma}(\alpha,\beta)$ represents a Gamma distribution with shape parameter $\alpha$ and rate parameter $\beta$, $\text{half-Normal}(\mu,\sigma^2)$ denotes a half-Normal distribution with location parameter $\mu$ and scale parameter $\sigma$, and $\Sigma_{g,ii}$ denotes the $i$th diagonal entry of $\Sigma_g$. For fixed effect and association parameters ($\beta_{gi}$, $\gamma_{gi}$, $\phi_{g1}$ and $\alpha_g$), the priors are weakly informative, similar to those used in standard JMs. We use a slightly more informative prior for the fixed effect parameters $\psi_{gi}$ in the class membership submodel to prevent the resulting mixture weights from getting too close to the boundary of the simplex. For the Weibull shape parameter $\phi_{g0}$, the Gamma prior we use has a flat, unimodal shape and right skewness, covering plausible values in survival analysis. For the variance parameters, the prior hyperparameters are partly informed from an empirical Bayes perspective. For $\sigma^{2}_{g}$, the scale parameter in the half-Normal prior is chosen to be close to the MLE results obtained from fitting a LMM to the longitudinal data. The shape and rate parameters in the Gamma prior for $\Sigma_{g,ii}$ are chosen to be $1.5$ (for $G>1$), which is found to sufficiently cover plausible values of the random effects variance (for $G=2,3,4$). When fitting the model with $G=1$, i.e., no latent class, we use the same prior settings as in \eqref{eq:priorsim}, except that for $\Sigma_{g,ii}$, we use an inverse gamma prior with shape and scale parameters set to $0.01$, which is a common weakly informative choice for the standard JM.

To perform posterior sampling, as described in Section \ref{subsec:mcmc}, we used the Stan modelling language via the R interface provided by the {\tt rstan} R package \citep{rstan}, which provides a convenient implementation of the NUTS. In our simulations, we used the default settings for the optional tuning parameters in NUTS, as they generally worked well and achieved a good balance between sampling and computational efficiency. However, in practice, for complex models where divergent transitions or stability issues arise, these default settings can be adjusted. For example, the target acceptance rate (referred to as $\text{adapt\_delta}$) can be increased from the default value of 0.8 to a larger value (such as 0.9 or higher) to improve the robustness of NUTS, although higher values may substantially increase sampling time. To implement the parallel sampling scheme as described in Section \ref{subsec:multimodality}, we used $M=6$ parallel chains, each run for $5500$ iterations, with the first $2500$ iterations discarded as burn-in. For each chain, the $3000$ post-burn-in samples were further thinned by an interval of 3. These settings were found to be sufficient for our simulation; however, in more challenging scenarios, more chains, a longer burn-in period, or a larger thinning interval may be required. To estimate the weights for each chain using the truncated harmonic mean estimator, we used $\beta = 0.6$ (but the results seem generally robust for different values of this parameter). All computations were performed on the Cambridge Service for Data Driven Discovery (CSD3) High-Performance Computing (HPC) system using the Ice Lake CPUs.

\subsection{Results}

Table \ref{tab:sim_modelchoice} summarizes the performance of LOOIC and WAIC in model selection under simulation Setting I, with results obtained using the procedure described in Section \ref{subsec:modelchoice}. LOOIC consistently outperforms WAIC, and as the significance level becomes more stringent, the performance of both criteria improves. When compared to the results reported in \citet{andrinopoulou2020} (see Table 3 therein for details), which shows proportions of the true number of classes selected at $30\%$, $49\%$ and $72\%$ for scenarios 1, 2 and 3 respectively, conditional on the best-performed tuning parameter, both LOOIC (at all significance levels) and WAIC (at $\alpha=0.1\%$) demonstrate superior accuracy across all three scenarios. Figures \ref{fig:sim_looicplots} and \ref{fig:sim_waicplots} in the supplementary material displays the results of LOOIC and WAIC (on the deviance scale, where lower values indicate better performance) across replications of the data. It is interesting to note that both criteria exhibit a general pattern where the estimates stabilise once the number of latent classes reaches the true value. It also becomes clear that there is a potential risk of overfitting if one simply selects the model with the lowest value. Supplemental Tables \ref{tab:sim_m1s3} and \ref{tab:sim_m2s1} (under Model 1) present the estimation results conditional on $G=3$ and $G=2$, respectively. Our proposed algorithm demonstrates satisfactory performance in both point and interval estimation. With regard to classification accuracy, defined as the proportion of correctly classified subjects based on the maximum {\it a posteriori} (MAP) decision rule, and with class allocation probabilities computed as described in Section \ref{subsec:mcmc}, we achieve consistently high accuracy. The median and interquartile range summarized from 200 replications are $98.8\% \ (98.6\%, 99.0\%)$ for $G=2$ and $99.4\% \ (99.2\%, 99.6\%)$ for $G=3$, respectively.

\begin{table}[h!]
\centering
\begin{tabular}{ccccc}
\toprule
\multirow{2}{*}{\textbf{Scenario}} & \multirow{2}{*}{\textbf{Criteria}} & \multicolumn{3}{c}{\textbf{True number of classes (\%)}} \\ 
\cmidrule(lr){3-5}
 &  & $z=1.65\ (\alpha=5\%)$ & $z=2.33\ (\alpha=1\%)$ & $z=3.09\ (\alpha=0.1\%)$ \\ 
\midrule
\multirow{2}{*}{\parbox{4cm}{\textbf{Scenario 1 (G = 1)}}} 
 & LOOIC  & 72.0 & 82.8 & 89.8 \\ 
 & WAIC   & 33.8 & 65.6 & 84.7 \\ 
\cmidrule(lr){2-5}
\multirow{2}{*}{\parbox{4cm}{\textbf{Scenario 2 (G = 2)}}} 
& LOOIC   & 67.0 & 80.9 & 90.4 \\ 
& WAIC  & 23.4 & 47.3 & 76.1 \\ 
\cmidrule(lr){2-5}
\multirow{2}{*}{\parbox{4cm}{\textbf{Scenario 3 (G = 3)}}} 
& LOOIC  & 77.7 & 83.8 & 88.8\\ 
& WAIC   & 45.2 & 68.0 & 85.8 \\ 
\bottomrule
\end{tabular}
\caption{Performance of LOOIC and WAIC under simulation Setting I.
Model comparison and selection were performed as described in Section \ref{subsec:modelchoice} of the main paper. Three different critical values for the z-score, corresponding to one-tailed tests with significance levels of $5\%$, $1\%$, and $0.1\%$, were considered. For each significance level, the true number of classes ($\%$) was computed as the proportion of times the correct number of classes was selected, based on 157, 188, and 197 replications of data for Scenarios 1, 2, and 3, respectively. Not all of the 200 replications were used because, when fitting models with $G$ larger than the true number of classes, sampling may have broken down due to class collapse, and the results from those replications were excluded. }
\label{tab:sim_modelchoice}
\end{table}

\begin{table}[h!]
\centering
\begin{tabular}{cccccccc}
\toprule
\multirow{2}{*}{\textbf{Parameter}} & \multirow{2}{*}{\textbf{True Value}} & \multicolumn{3}{c}{\textbf{Model 1}} & \multicolumn{3}{c}{\textbf{Model 2}} \\
\cmidrule(lr){3-5} \cmidrule(lr){6-8}
& & Bias & SD & Coverage (\%) & Bias & SD & Coverage (\%) \\
\midrule
$\beta_{1,0}$ & 8.03 & -0.002 & 0.142 & 94.0  & -0.002 & 0.141 & 94.5 \\
$\beta_{1,1}$ & -0.16 &  0.001 & 0.011 & 94.0  &  0.000 & 0.011 & 95.5 \\
$\beta_{1,2}$ & -5.86 & -0.100 & 0.143 & 85.5  & -0.002 & 0.143 & 94.5 \\
$\beta_{2,0}$ & -8.03 &  0.009 & 0.034 & 94.5  &  0.002 & 0.034 & 94.0 \\
$\beta_{2,1}$ & 0.46 & -0.062 & 0.072 & 83.5  &  0.002 & 0.071 & 94.5 \\
$\beta_{2,2}$ & 12.2 & -0.141 & 0.082 & 53.0  & -0.004 & 0.072 & 94.0 \\
$\gamma_{1,0}$ & -4.85 & -0.336 & 0.473 & 85.5  & -0.031 & 0.413 & 94.0 \\
$\gamma_{1,1}$ & -0.02 & -0.002 & 0.007 & 91.0  & -0.001 & 0.006 & 92.5 \\
$\gamma_{2,0}$ & -4.85 & -0.172 & 0.276 & 87.5  & -0.026 & 0.283 & 94.0 \\
$\gamma_{2,1}$ & 0.09 &  0.002 & 0.005 & 92.5  &  0.000 & 0.005 & 94.0 \\
$\alpha_{1}$ & 0.38 &  0.021 & 0.037 & 90.5  &  0.001 & 0.034 & 93.5 \\
$\alpha_{2}$ & 0.08 & -0.008 & 0.010 & 82.0  &  0.000 & 0.009 & 94.5 \\
$\xi_{1}$ & 1.8 &  0.121 & 0.129 & 85.5  &  0.023 & 0.117 & 97.5 \\
$\xi_{2}$ & 1.4 &  0.006 & 0.058 & 93.5  &  0.011 & 0.058 & 93.5 \\
$\sigma^{2}_{1}$ & 0.4761 &  0.000 & 0.015 & 94.0  &  0.001 & 0.015 & 94.0 \\
$\sigma^{2}_{2}$ & 0.4761 & -0.005 & 0.023 & 96.0  & -0.007 & 0.023 & 96.0 \\
$\Sigma_{1,11}$ & 0.87 & -0.083 & 0.066 & 80.0  & -0.001 & 0.072 & 94.5 \\
$\Sigma_{1,22}$ & 0.02 &  0.000 & 0.002 & 95.5  &  0.001 & 0.002 & 96.0 \\
$\Sigma_{2,11}$ & 0.02 &  0.031 & 0.025 & 74.5  &  0.015 & 0.019 & 92.5 \\
$\Sigma_{2,22}$ & 0.91 & -0.016 & 0.096 & 93.5  & -0.004 & 0.098 & 95.0 \\
\multicolumn{2}{c}{} & \multicolumn{3}{c}{Classification accuracy} & \multicolumn{3}{c}{Classification accuracy} \\
\multicolumn{2}{c}{} & \multicolumn{3}{c}{97.4 (97.0, 97.9)} & \multicolumn{3}{c}{99.1 (98.9, 99.3)} \\
\bottomrule
\end{tabular}
\caption{Results for simulation Setting II (Scenario 3). For each parameter, bias and standard deviation (SD) are evaluated based on the posterior mean obtained from 200 data replications, and coverage is the proportion of times the 95\% credible interval contains the true parameter value across the 200 replications. Classification accuracy is summarized using the median and interquartile range (shown in brackets), with accuracy for each replication calculated as the proportion of correctly classified subjects.
Model 1: model with homogeneous mixture weights (ignoring covariate effects on the class membership). Model 2: model with covariate dependent mixture weights (use gender and age only).}
\label{tab:sim_m2s3}
\end{table}

\begin{table}[h!]
\centering
\begin{tabular}{cccccccc}
\toprule
\multirow{2}{*}{\textbf{Parameter}} & \multirow{2}{*}{\textbf{True Value}} & \multicolumn{3}{c}{\textbf{Model 1}} & \multicolumn{3}{c}{\textbf{Model 2}} \\
\cmidrule(lr){3-5} \cmidrule(lr){6-8}
& & Bias & SD & Coverage (\%) & Bias & SD & Coverage (\%) \\
\midrule
$\beta_{1,0}$    & 8.03   & -0.023 & 0.127 & 94.5  & -0.025 & 0.127 & 96.0 \\
$\beta_{1,1}$    & -0.16  &  0.001 & 0.010 & 95.5  &  0.000 & 0.010 & 95.5 \\
$\beta_{1,2}$    & -5.86  & -0.071 & 0.144 & 87.5  &  0.026 & 0.144 & 93.5 \\
$\beta_{2,0}$    & -8.03  &  0.011 & 0.033 & 93.5  &  0.004 & 0.032 & 93.5 \\
$\beta_{2,1}$    & 0.46   & -0.061 & 0.076 & 81.5  &  0.000 & 0.079 & 89.0 \\
$\beta_{2,2}$    & 12.2   & -0.134 & 0.083 & 53.5  & -0.011 & 0.072 & 96.5 \\
$\gamma_{1,0}$     & -4.85  & -0.370 & 0.435 & 86.5  & -0.062 & 0.397 & 93.0 \\
$\gamma_{1,1}$     & -0.02  & -0.001 & 0.006 & 95.0  & -0.001 & 0.006 & 96.0 \\
$\gamma_{2,0}$     & -4.85  & -0.176 & 0.261 & 91.5  & -0.051 & 0.267 & 95.0 \\
$\gamma_{2,1}$     &  0.09  &  0.002 & 0.004 & 93.5  &  0.001 & 0.005 & 94.5 \\
$\alpha_{1}$      &  0.38  &  0.025 & 0.034 & 88.5  &  0.004 & 0.032 & 96.0 \\
$\alpha_{2}$      &  0.08  & -0.007 & 0.009 & 86.0  &  0.000 & 0.009 & 97.5 \\
$\xi_{1}$     &  1.8   &  0.128 & 0.132 & 82.0  &  0.031 & 0.123 & 95.5 \\
$\xi_{2}$     &  1.4   &  0.005 & 0.051 & 94.0  &  0.010 & 0.052 & 94.0 \\
$\sigma^{2}_{1}$ & 0.4761 &  0.000 & 0.014 & 93.5  &  0.001 & 0.014 & 94.5 \\
$\sigma^{2}_{2}$ & 0.4761 & -0.004 & 0.026 & 95.0  & -0.006 & 0.025 & 94.0 \\
$\Sigma_{1,11}$         &  0.87  & -0.064 & 0.071 & 84.5  &  0.012 & 0.076 & 94.5 \\
$\Sigma_{1,22}$         &  0.02  &  0.000 & 0.002 & 93.5  &  0.001 & 0.002 & 94.5 \\
$\Sigma_{2,11}$         &  0.02  &  0.034 & 0.026 & 73.0  &  0.018 & 0.019 & 88.5 \\
$\Sigma_{2,22}$         &  0.91  & -0.009 & 0.086 & 97.5  &  0.002 & 0.087 & 98.0 \\
\multicolumn{2}{c}{} & \multicolumn{3}{c}{Classification accuracy} & \multicolumn{3}{c}{Classification accuracy} \\
\multicolumn{2}{c}{} & \multicolumn{3}{c}{97.5 (97.1, 98.0)} & \multicolumn{3}{c}{99.0 (98.8, 99.3)} \\
\bottomrule
\end{tabular}
\caption{Results for simulation Setting II (Scenario 4). The settings are the same as those in Table \ref{tab:sim_m2s3}.}
\label{tab:sim_m2s4}
\end{table}

For simulation Setting II, Tables \ref{tab:sim_m2s3} and \ref{tab:sim_m2s4} summarise the estimation results for Scenarios 3 and 4, respectively, while results for Scenarios 1 and 2 are presented in Supplemental Tables \ref{tab:sim_m2s1} and \ref{tab:sim_m2s2}. Model 1, which ignores covariates in class membership submodel, can incur significant bias and much lower coverage for certain parameters when the true class allocation depends on covariates that are also related to the longitudinal and time-to-event processes (see Tables \ref{tab:sim_m2s3} and \ref{tab:sim_m2s4}). Model 2, which uses the same covariates as those considered in the class-specific JM, provides consistently satisfactory performance regardless of the ground truth, offering much better estimation accuracy in Scenarios 3 and 4 (see Tables \ref{tab:sim_m2s3} and \ref{tab:sim_m2s4}) and similar accuracy in Scenarios 1 and 2 (see Supplemental Tables \ref{tab:sim_m2s1} and \ref{tab:sim_m2s2}) compared to Model 1. Note that the classification performance appears to be minimally impacted by these specifications. Although not shown here, we conducted additional analyses under various settings. We considered an alternative version of Model 1, where the softmax parametrisation for the class membership submodel is replaced by a Dirichlet model on the homogeneous mixture weights, with the concentration parameter set to 1. The results are nearly identical, and the key patterns remaining unchanged. When $G$ is not fixed, we found that ignoring relevant covariates in the membership submodel also impacts the selection of the number of classes. In particular, we found that this omission can be compensated for by selecting additional classes. Clearly, the extent of the bias incurred and the impact on inference depend on the strength of the association between the covariates and the class allocation and longitudinal processes. In general, we would expect the impact to increase as the strength of the association grows.

\section{Application to the PAQUID dataset} \label{sec:appl}

We illustrate our proposed method using the PAQUID dataset available from the {\tt lcmm} R package, which consists of a random subsample of 500 subjects from the prospective cohort study of the same name, aimed at investigating the relationship between risk factors and cognitive and functional aging diseases in the elderly population in France. The study participants were aged 65 and over at entry and were drawn from two regions in southwestern France. Information on baseline socio-demographic variables, cognitive performance, and health and medical history were recorded at an individual level over a maximum follow up period of 20 years. For more background information on the original data, we refer to \citet{2017lcmm} and the references therein.
\citet{2017lcmm} used the basic JLCM to examine the trajectory of repeated measures of a normalized version of the Mini-Mental State Examination score (normMMSE) with age and the associated risk of dementia. For this analysis, 1 subject was removed due to the event of interest occurring earlier than the entry time ($n=499$). The number of longitudinal measurements per individual had a median of 4 and an interquartile range of $(2,7)$, and $74.3\%$ of subjects had censored event times. Using the basic JLCM, model selection on the number of latent classes was performed based on the BIC, and $G=4$ latent classes was suggested. Interestingly, with $G=4$, the hypothesis testing performed therein indicated that the conditional independence of the longitudinal and time-to-event outcomes, given latent class, was not supported by the data.

Here, motivated by the analysis conducted in \citet{2017lcmm}, we use the shared parameter JLCM to re-analyse the same data to explore any changes in the latent class structure and to investigate the relationship between normMMSE and the risk of dementia within a latent class, which remains unexplored in earlier works. The basic setup for the longitudinal and time-to-event submodels mostly follows the JLCM used in \citet{2017lcmm}. To simplify prior specification and improve sampling efficiency, we further standardised the normMMSE (which ranges from 0 to 100) to have a mean of 0 and a standard deviation of 1. The longitudinal trajectory for the standardized normMMSE, conditional on the class membership $k$, is modelled via a LMM as
\begin{equation}
    (y_{ij} \mid c_i = g) = \beta_{g0} + \beta_{g1} age65_{ij} + \beta_{g2} age65_{ij}^2 + b_{ig0} + b_{ig1} age65_{ij} + b_{ig2} age65_{ij}^2 + \beta_{g,CEP} CEP_i + \epsilon_{igj},
\end{equation}
where $age65_{ij} = (age_{ij} - 65) / 10$, with $age_{ij}$ being the age of individual $i$ at the $j$th visit, and $CEP_i$ is a binary variable for education, where 1 indicates the subject graduated with a primary school diploma, and 0 otherwise. We further assumed that $(b_{ig0},b_{ig1},b_{ig2})\sim N(0,\Sigma_g)$, with $\Sigma_g$ being diagonal, and that $\epsilon_{igj}\sim N(0,\sigma_{g}^{2})$. 
For the class-specific time-to-event submodel, we used the same time scale (i.e., $age65_{ij}$) as for the longitudinal process, and we link the time-to-event process with the longitudinal process using a current value association. The hazard function is specified as
\begin{equation}
    h_i(t \mid c_i = g) = h_{0g}(t) \exp\left( \gamma_{g,CEP} CEP_i + \gamma_{g,male} male_i + \gamma_{g,CEP \times male} (CEP_i \times male_i) + \alpha_g \mu_{ig}(t) \right),
\end{equation}
where $h_{0g}(t) = \phi_{g0} t^{\phi_{g0} - 1} \exp(\phi_{g1})$, $male_i$ is a binary variable, with 1 for male and 0 for female, and $\mu_{ig}(t)$ is the conditional mean of $y_{i}(t)$ given $c_i=g$. The interaction term was not considered in the previous JLCM, but we include it here to assess whether an interaction effect between gender and education occurs. We also note that \citet{2017lcmm} did not incorporate covariates into the class membership submodel. In our analysis, however, motivated by our findings from the simulation study, we chose to include both $CEP_i$ and $male_i$ to guard against the possibility of incurring any bias on the estimation results. The membership submodel is therefore specified as 
\begin{equation}
\pi_{ig} = P(c_{i}=g) = \frac{\exp(\psi_{g0}+\psi_{g,CEP}CEP_i +\psi_{g,male}male_i)}{\sum_{k=1}^{G}\exp(\psi_{k0}+\psi_{k,CEP}CEP_i +\psi_{k,male}male_i)}, \quad g=1,\ldots,G,
\end{equation}
where $\psi_{G0}$, $\psi_{G,CEP}$ and $\psi_{G,male}$ are set to zeros.
To complete the Bayesian model, we choose the following set of prior distributions:
\begin{equation}\label{eq:priorapp}
\begin{gathered}
   \beta_{g\cdot},  \psi_{g\cdot},  \sim N(0,2^2), \quad \gamma_{g\cdot}, \phi_{g1}, \alpha_g \sim N(0,3^2), \quad \phi_{g0} \sim \text{Gamma}(2,0.5),\\
   \sigma^{2}_{g}\sim \text{half-Normal}(0,0.2^2), \quad \Sigma_{g,ii}\sim \text{Gamma}(1.5,1.5). 
\end{gathered}
\end{equation}
Note that we used generally more informative normal priors for the fixed effect parameters than in the simulation study, partly due to the smaller sample size here, and we also accounted for the fact that the longitudinal data had been standardized. The hyperparameters for the priors on the variance parameters were informed by the MLE results obtained from the {\tt lcmm} R package, as was done in the simulation study.

\begin{figure}[!h]
    \centering
    \includegraphics[width=0.45\textwidth]{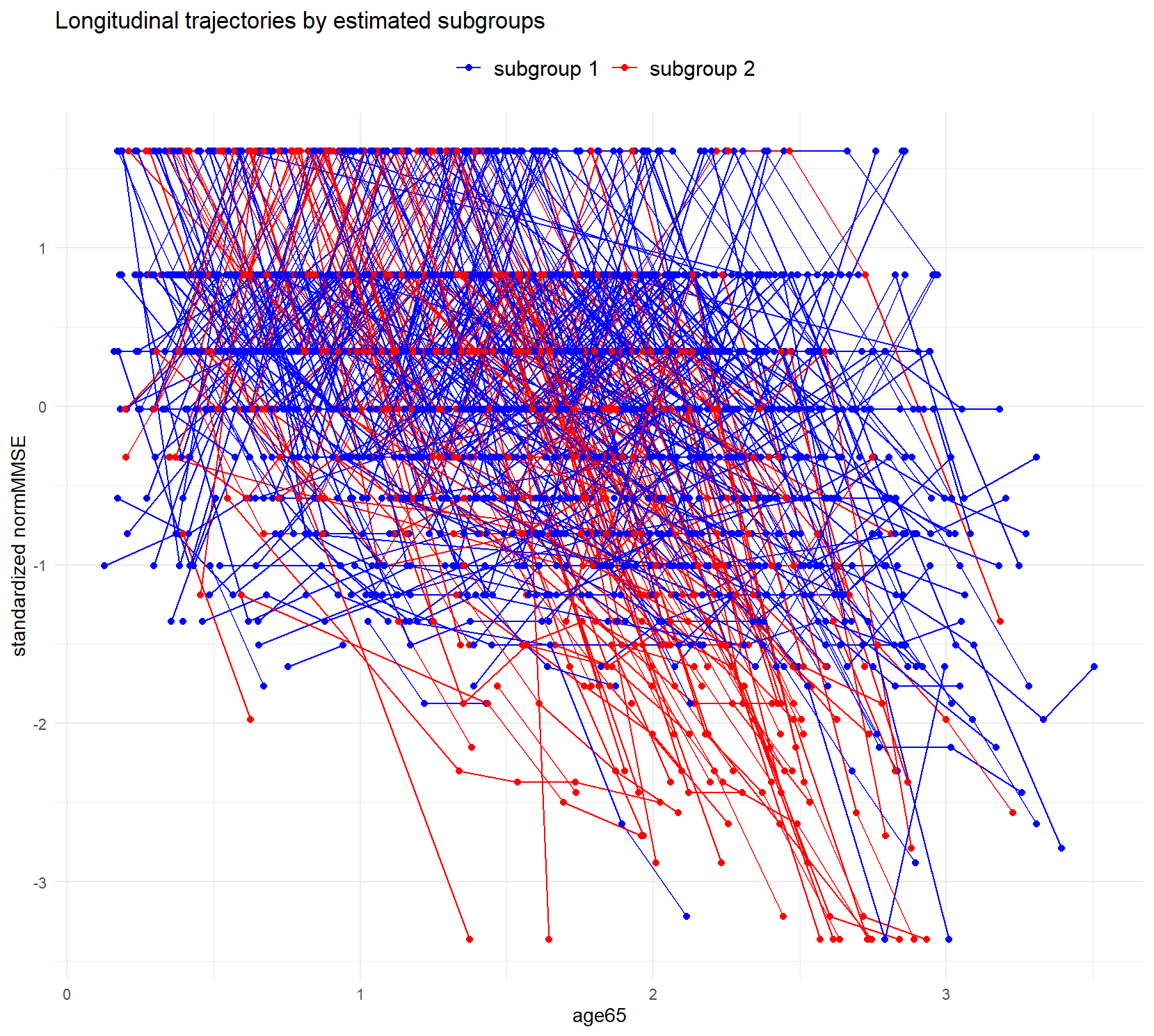}\includegraphics[width=0.45\textwidth]{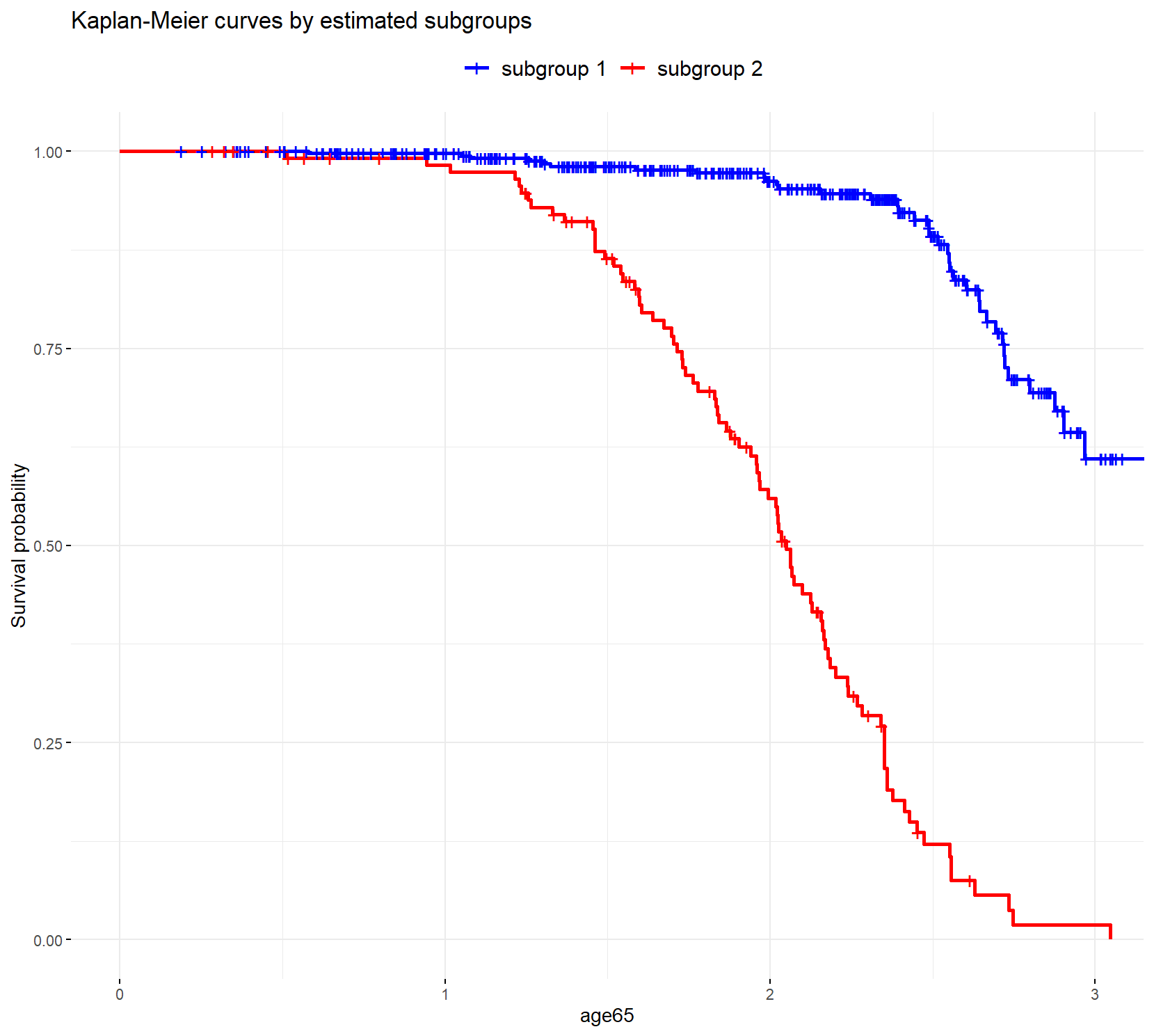}
    \includegraphics[width=0.45\textwidth]{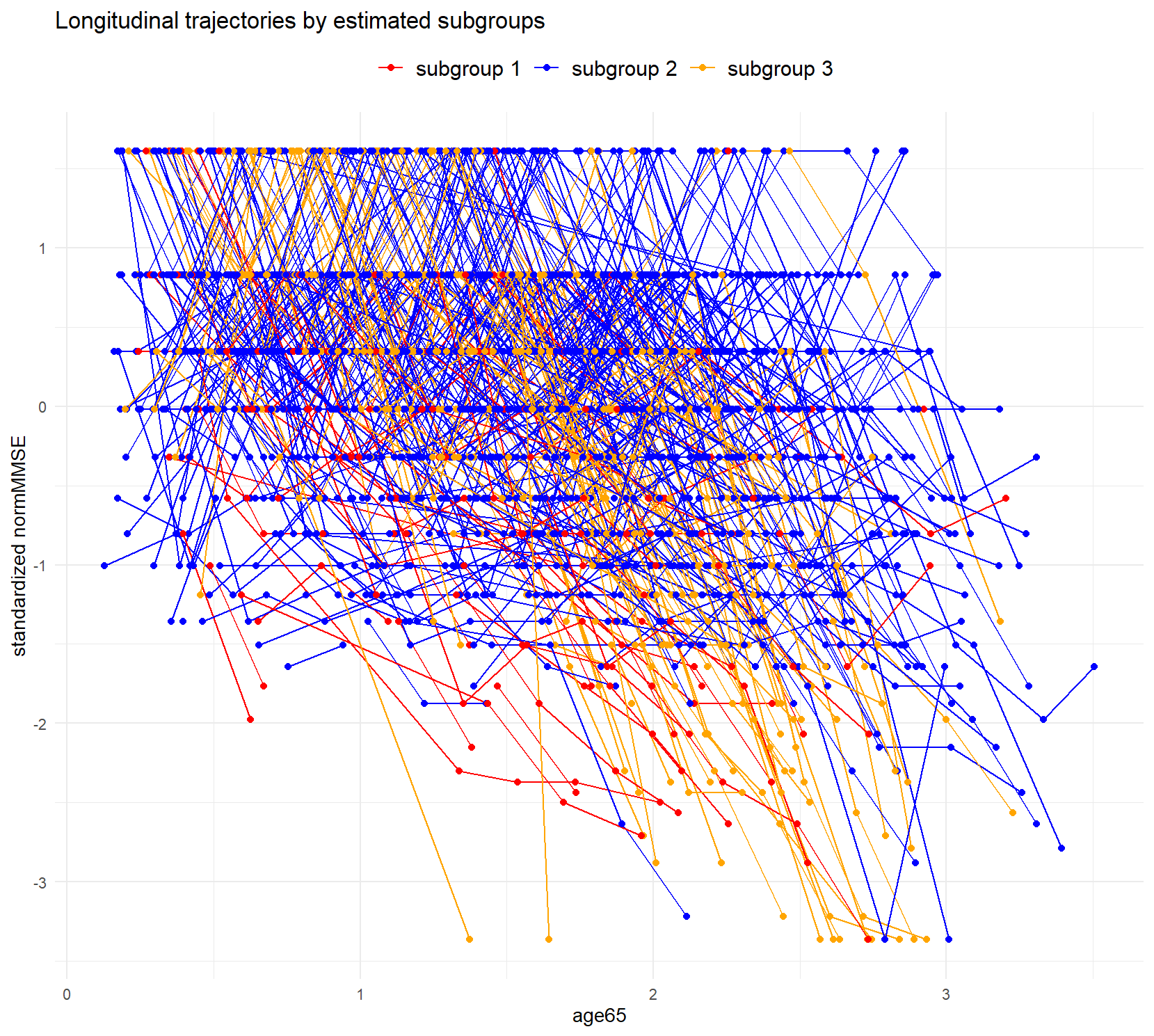}\includegraphics[width=0.45\textwidth]{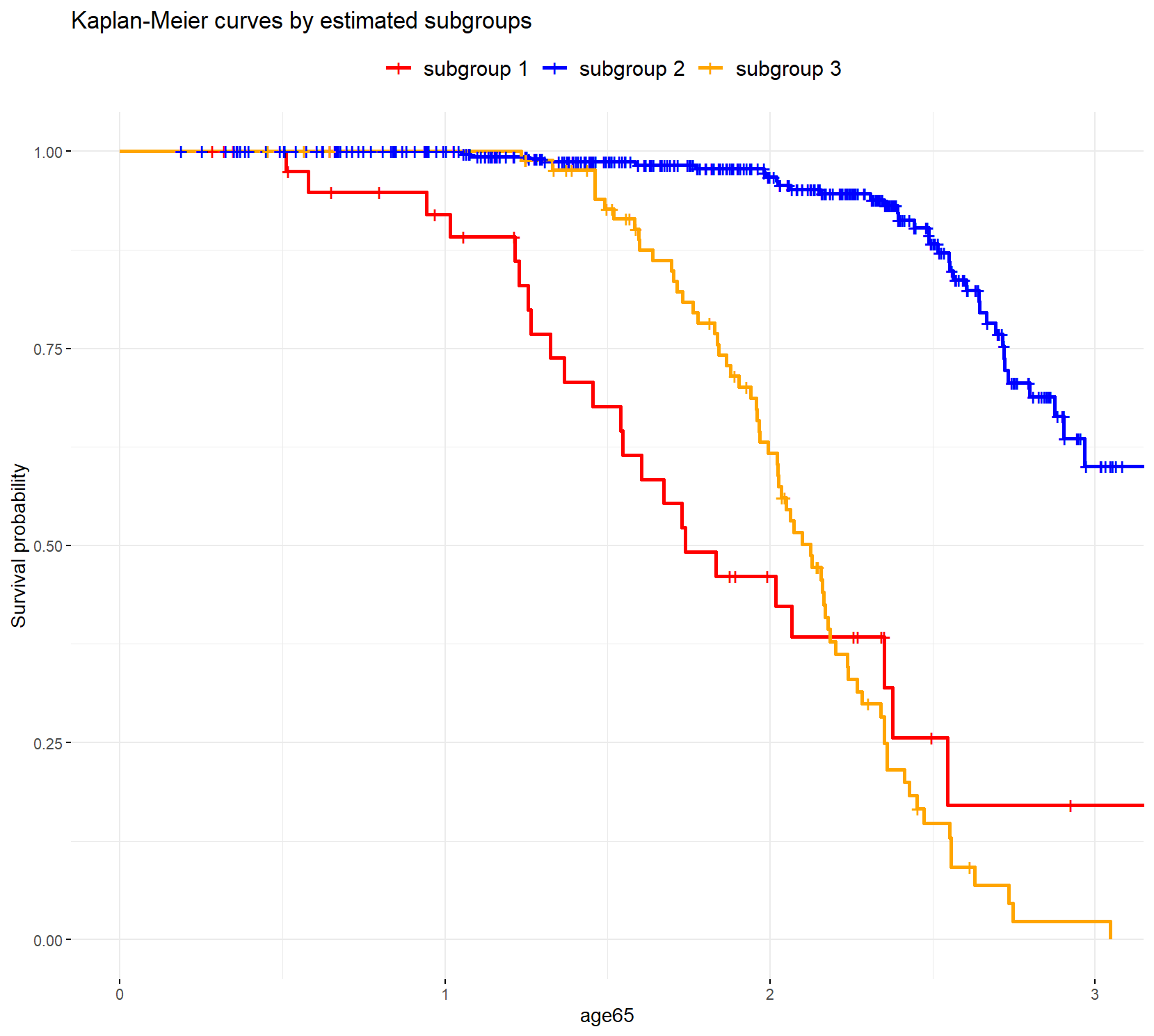}
    \caption{Trajectories of standardized normMMSE and Kaplan-Meier curves by estimated subgroups (indicated by colours). The top panel displays results for $G=2$, and the bottom panel for $G=3$.}
    \label{fig:app_JLCMplots}
\end{figure}

We first examine the number of latent classes using the approach described in Section \ref{subsec:modelchoice}, considering candidate values of $G$ ranging from one to four due to the relatively small sample size. Posterior inference was based on $M=6$ parallel chains, each run for $12000$ iterations (for $G=1, 2, 3$) or $14000$ iterations (for $G=4$), with the first $4000$ (for $G=1, 2, 3$) or $6000$ (for $G=4$) iterations discarded as burn-in, based on pilot runs and convergence diagnostics. For each chain, the $8000$ post burn-in samples were further thinned by an interval of 4. The other estimation settings are the same as in the simulation study.
Table \ref{tab:app_modelcompare} shows the results of the estimated LOOIC and WAIC for different values of $G$ (on the deviance scale), as well as the associated z-score for testing the difference in LOOIC or WAIC between models with $G=k$ and $G=k+1$. The largest drop in both LOOIC and WAIC occurs when increasing $G$ from one to two, with the z-scores associated with the difference in LOOIC and WAIC both greater than 5.6 (and p-values effectively zero), indicating strong support for the presence of a latent class structure. $G=4$ achieves the smallest LOOIC and WAIC; however, the improvement it offers over $G=3$ is minimal, as indicated by the associated z-scores (and p-values) for LOOIC and WAIC. Additionally, when examining the estimated latent class structure with $G=4$, an extra small cluster is generated with only three individuals assigned to it. Therefore, we may be skeptical in considering this as a meaningful `subgroup'. Comparing models with two and three latent classes, there seems to be moderate evidence for the difference in LOOIC and WAIC, as indicated by the associated z-scores and p-values. Figure \ref{fig:app_JLCMplots} displays the class-specific longitudinal trajectories and Kaplan-Meier survival curves based on the estimated models with $G=2$ (top panel) and $G=3$ (bottom panel). Class membership for each individual was assigned based on the maximum posterior class probability, as done in the simulation study. It appears that class 2 in the model with $G=2$, which is associated with more significant declines in normMMSE and a more rapid progression of dementia, is further split into two subclasses in the model with $G=3$ (labelled as subgroups 1 and 3). This visual impression is confirmed by comparing the class allocation results obtained from $G=2$ and $G=3$. Depending on the desired level of detail, either model could be of interest. Here, we focus on $G=2$ to capture the major subgroup structures within the study cohort. Based on the estimated class allocation, $75.8\%$ (377/499) of individuals are classified into class 1, the relatively lower-risk group, while the remaining belong to class 2. Table \ref{tab:app_JLCMresults} shows the posterior means and $95\%$ credible intervals for key parameters of interest in the model with $G=2$. The male gender shows a significant association with class membership ($\psi_{1,male}$), where, prior to observing any longitudinal data, being male increases the chance of being classified into the relatively lower-risk group. Additionally, there is a strong association between normMMSE and dementia, conditional on latent class membership, as indicated by the magnitude of $\alpha_k$ and the fact that the associated credible intervals exclude zero. The strength of the association is similar across both subgroups. The negative relationship is expected, as higher normMMSE generally reflects better cognitive function, which is anticipated to be associated with a lower risk of cognitive diseases such as dementia. The log-hazard ratio for education, $\gamma_{g,CEP}$, is estimated to be positive. While this may seem contradictory to earlier results in \citet{2017lcmm} where CEP had a protective effect on dementia, it is important to note that here the effect of CEP is mediated through the longitudinal marker (MMSE) via $\mu_i(t)$. The total effect of CEP on the risk of dementia, given by $\gamma_{g,CEP} + \alpha_g \times \beta_{g,CEP} + \gamma_{g,CEP \times male} \times male_i$, remains negative. Another observation is that the two subgroups exhibit different levels of within-group variability in normMMSE (measured via $\sigma_g^{2}$), with the higher-risk group, class 2, showing greater within-class variability. This aligns with the clinical literature, which links neurodegenerative disorders to increased intraindividual variability in cognitive performance metrics \citep{hultsch2000intraindividual,gorus2008reaction}.

\begin{table}[ht]
\centering
\begin{tabular}{ccccccc}
\toprule
 & \textbf{LOOIC} & $\bm {Z_{LOOIC}}$ & $\bm {P_{LOOIC}}$ & \textbf{WAIC} & ${\bm Z_{WAIC}}$ & $\bm {P_{WAIC}}$\\ 
\hline
\textbf{G = 1} & 4299 & 5.62 & 0 & 4068 & 6.97 & 0 \\ 
\textbf{G = 2} & 4118 & 2.82 & 0.002 & 3850 & 2.66 & 0.004 \\ 
\textbf{G = 3} & 4063 & 1.07 & 0.142 & 3802 &  0.63 & 0.264 \\ 
\textbf{G = 4} & 4033 & - & - & 3781 & - & - \\ 
\bottomrule
\end{tabular}
\caption{Results for comparing models with candidate values of $G$. For each $G=k$, LOOIC and WAIC (on the deviance scale) are computed as described in Sections \ref{subsec:modelchoice} and \ref{sec:appl}. The $Z_{LOOIC}$ and $Z_{WAIC}$ represent the z-scores associated with the differences in LOOIC and WAIC, respectively, comparing models with $G=k$ and $G=k+1$. ${P_{LOOIC}}$ and ${P_{WAIC}}$ are the one-tailed p-values associated with $Z_{LOOIC}$ and $Z_{WAIC}$, respectively.  }
\label{tab:app_modelcompare}
\end{table}

\begin{table}[!h]
\centering
\begin{tabular}{lccc}
\toprule
& \multirow{2}{*}{\textbf{Parameters}} & \multicolumn{2}{c}{\textbf{Estimates (Posterior Mean (95\% CI))}} \\
\cmidrule(lr){3-4}
& & Class 1 & Class 2 \\
\midrule

\multirow{2}{*}{Longitudinal submodel} & $\beta_{g,CEP}$ & 0.851 (0.670, 1.040) & 0.365 (-0.013, 0.705) \\
& $\sigma^{2}_{g}$ & 0.223 (0.191, 0.252) & 0.316 (0.268, 0.370) \\

\midrule

\multirow{4}{*}{Time-to-event submodel} & $\alpha_{g}$ & -3.697 (-5.308, -2.388) & -3.602 (-4.862, -2.581) \\
& $\gamma_{g,CEP}$ & 1.155 (-0.078, 2.466) & 0.460 (-0.506, 1.334) \\
& $\gamma_{g,male}$ & 0.424 (-1.053, 1.890) & 0.144 (-1.581, 1.666) \\
& $\gamma_{g,CEP\times male}$ & 0.921 (-1.029, 2.949) & -0.016 (-1.810, 1.943) \\

\midrule

\multirow{3}{*}{Membership submodel} & $\psi_{10}$ & \multicolumn{2}{c}{0.502 (-0.244, 1.315)} \\
& $\psi_{1,CEP}$ & \multicolumn{2}{c}{-0.245 (-1.016, 0.507)} \\
& $\psi_{1,male}$ & \multicolumn{2}{c}{ 0.904 (0.214, 1.581)} \\

\bottomrule
\end{tabular}
\caption{Estimation results for the shared parameter JLCM with $G=2$.}
\label{tab:app_JLCMresults}
\end{table}

\section{Discussion} \label{sec:disc}

In this paper, we focus on shared parameter JLCMs, which provide an important extension of the standard JM for study populations of a heterogeneous nature. We propose a new Bayesian inferential framework to effectively tackle the computational challenges faced by existing Bayesian methods. To enhance sampling efficiency, we adopt a strategy that enables the use of state-of-the-art MCMC methods. We effectively address potential multimodality in the posterior through a parallel sampling scheme, leveraging parallel computing power. Model selection for the number of latent classes can be guided by predictive-based criteria such as LOOIC, which are conveniently estimated from MCMC output. Through a simulation study, we demonstrate that the proposed method provides superior estimation accuracy compared to the existing approach. The feasibility of our method is further illustrated via an application to the PAQUID data, where we explored the underlying latent class structure and obtained insights into the association between the cognitive measure MMSE and the risk of dementia within each latent class.

Our results bring important insights to the practical implementation of such models. One important aspect regards the prior specification, which has received very little discussion in the context of shared parameter JLCMs. Unlike the standard JM, prior specifications for JLCMs require additional care. It is desirable to use relatively informative priors that exploit available knowledge, perhaps in conjunction with empirical Bayes strategies, which can help with model identifiability and the convergence of the algorithm. Sensitivity analysis may be performed to examine the impact in specific application contexts. Regarding selection on the number of latent classes, our results suggest that predictive-based criteria such as LOOIC can be informative, though the hypothesis testing procedure introduced in Section \ref{subsec:modelchoice} (i.e., testing whether the improvement offered by the more complex model is truly significant) is necessary to guard against the risk of overfitting. In particular, a more conservative approach (i.e., using a relatively stringent significance level) seems necessary. We also found that including relevant covariates in the class membership model is important. While ignoring these covariates has minimal impact on classification accuracy, bias arises when they are related to the longitudinal and time-to-event processes. Therefore, it seems advisable to include the same set of covariates considered in class-specific JMs to mitigate potential bias.

From the modelling perspective, it is possible to extend the algorithm considered here to alternative formulations of JLCMs, depending on the application context. For instance, when the focus is on prediction and multivariate longitudinal markers are present, we could replace the shared parameter structure with a two-stage approach, where summaries from the raw longitudinal data are first extracted and treated as time-varying covariates in the time-to-event submodel \citep{alvares2023}. A specific formulation under a discrete hazard framework using a frequentist approach has been considered in \citet{nguyen2023flash}. A Bayesian formulation under a more generic hazard setting may provide advantages in terms of prior regularization, prediction, and associated uncertainty quantification. From the algorithmic perspective, it would be interesting to explore alternative approximate Bayesian inference methods that are potentially more scalable in complex, large-data settings. The MCMC-based method presented here can serve as a gold standard benchmark for evaluating these approximate algorithms. One option, motivated by the recent success of INLA for the standard JMs \citep{rustand2024fast, alvares2024}, is the possibility of adapting it to the mixture setting by combining MCMC with INLA \citep{gomez2018markov}. In this approach, we sample from the marginal posterior of the latent class allocation variables using MCMC, and conditional on the sampled class allocation, INLA is used to fit the class-specific JM. Some preliminary experiments (not reported here) suggest that the efficiency was extremely low due to the poor acceptance rate of the MCMC step and stability issues with INLA, which arise from the need to refit the model a large number of times. A more efficient MCMC proposal and a more stable or dedicated version of the INLA program warrant further development. Variational-based methods could offer another promising option. Black-box variational algorithms, such as those provided by Stan, are not feasible for such complex models. A customized development, possibly building on recent advances in variational inference algorithms for standard JMs \citep{sun2024penalized}, would be of interest.

% \section*{Data availability statement}

 \section*{Acknowledgements}

SC and MP were supported by the UK Medical Research Council (MRC) grant ``Looking beyond the mean: what within-person variability can tell us about dementia, cardiovascular disease and cystic fibrosis'' (MR/V020595/1). DA and JKB were supported by MRC Unit Programme MC\_UU\_00002/5.

\bibliographystyle{agsm}
\bibliography{refs}

@Misc{rstan,
    title = {{RStan}: {T}he {R} interface to {Stan}},
    author = {{Stan Development Team}},
    note = {R package version 2.32.6},
    year = {2024},
    url = {https://mc-stan.org/}
  }

@Misc{loo,
    title = {loo: {E}fficient leave-one-out cross-validation and {WAIC} for
      {B}ayesian models},
    author = {Aki Vehtari and Jonah Gabry and Måns Magnusson and Yuling
      Yao and Paul-Christian Bürkner and Topi Paananen and Andrew
      Gelman},
    year = {2024},
    note = {R package version 2.8.0},
    url = {https://mc-stan.org/loo/},
  }

@book{rizopoulos2012joint,
  title={Joint models for longitudinal and time-to-event data: {W}ith applications in {R}},
  author={Rizopoulos, D.},
  year={2012},
  publisher={Chapman \& Hall/CRC}
}

@article{proust2014joint,
  title={Joint latent class models for longitudinal and time-to-event data: {A} review},
  author={Proust-Lima, C{\'e}cile and S{\'e}ne, Mb{\'e}ry and Taylor, Jeremy MG and Jacqmin-Gadda, H{\'e}l{\`e}ne},
  journal={Statistical Methods in Medical Research},
  volume={23},
  number={1},
  pages={74--90},
  year={2014}
}

@article{andrinopoulou2020,
  title={Integrating latent classes in the {B}ayesian shared parameter joint model of longitudinal and survival outcomes},
  author={Andrinopoulou, E. R. and Nasserinejad, K. and Szczesniak, R. and Rizopoulos, D.},
  journal={Statistical Methods in Medical Research},
  volume={29},
  number={11},
  pages={3294--3307},
  year={2020}
}

@article{vstrumbelj2024past,
  title={Past, present and future of software for {B}ayesian inference},
  author={{\v{S}}trumbelj, Erik and Bouchard-C{\^o}t{\'e}, Alexandre and Corander, Jukka and Gelman, Andrew and Rue, H{\aa}vard and Murray, Lawrence and Pesonen, Henri and Plummer, Martyn and Vehtari, Aki},
  journal={Statistical Science},
  volume={39},
  number={1},
  pages={46--61},
  year={2024}
}

@article{yao2022stacking,
  title={Stacking for non-mixing {B}ayesian computations: {T}he curse and blessing of multimodal posteriors},
  author={Yao, Yuling and Vehtari, Aki and Gelman, Andrew},
  journal={Journal of Machine Learning Research},
  volume={23},
  number={79},
  pages={1--45},
  year={2022}
}

@article{smith1992bayesian,
  title={Bayesian statistics without tears: {A} sampling--resampling perspective},
  author={Smith, Adrian FM and Gelfand, Alan E},
  journal={The American Statistician},
  volume={46},
  number={2},
  pages={84--88},
  year={1992}
}

@article{stephens2000dealing,
  title={Dealing with label switching in mixture models},
  author={Stephens, Matthew},
  journal={Journal of the Royal Statistical Society: Series B (Statistical Methodology)},
  volume={62},
  number={4},
  pages={795--809},
  year={2000}
}

@article{Robert2009,
    author = {Robert, C. P. and Wraith, D.},
    title = "{Computational methods for {B}ayesian model choice}",
    journal = {AIP Conference Proceedings},
    volume = {1193},
    number = {1},
    pages = {251-262},
    year = {2009},
    month = {12}
}

@article{chen2023bayesian,
  title={Bayesian spline-based hidden {M}arkov models with applications to actimetry data and sleep analysis},
  author={Chen, Sida and Finkenst{\"a}dt, B{\"a}rbel},
  journal={Journal of the American Statistical Association},
  pages={1--11},
  year={2023}
}

@article{vehtari2017practical,
  title={Practical {B}ayesian model evaluation using leave-one-out cross-validation and {WAIC}},
  author={Vehtari, Aki and Gelman, Andrew and Gabry, Jonah},
  journal={Statistics and Computing},
  volume={27},
  pages={1413--1432},
  year={2017}
}

@article{vehtari2021rank,
  title={Rank-normalization, folding, and localization: {A}n improved $\hat{R}$ for assessing convergence of {MCMC} (with discussion)},
  author={Vehtari, Aki and Gelman, Andrew and Simpson, Daniel and Carpenter, Bob and B{\"u}rkner, Paul-Christian},
  journal={Bayesian Analysis},
  volume={16},
  number={2},
  pages={667--718},
  year={2021}
}

@article{llorente2023marginal,
  title={Marginal likelihood computation for model selection and hypothesis testing: {A}n extensive review},
  author={Llorente, Fernando and Martino, Luca and Delgado, David and {Lopez-Santiago}, Javier},
  journal={SIAM review},
  volume={65},
  number={1},
  pages={3--58},
  year={2023},
  publisher={SIAM}
}

@article{2017lcmm,
 title={Estimation of extended mixed models using latent classes and latent processes: {T}he {R} package lcmm},
 volume={78},
 number={2},
 journal={Journal of Statistical Software},
 author={Proust-Lima, Cécile and Philipps, Viviane and Liquet, Benoit},
 year={2017},
 pages={1–56}
}

@article{liu2015joint,
  title={Joint latent class model of survival and longitudinal data: {A}n application to {CPCRA} study},
  author={Liu, Yue and Liu, Lei and Zhou, Jianhui},
  journal={Computational Statistics \& Data Analysis},
  volume={91},
  pages={40--50},
  year={2015}
}

@article{papageorgiou2019overview,
  title={An overview of joint modeling of time-to-event and longitudinal outcomes},
  author={Papageorgiou, Grigorios and Mauff, Katya and Tomer, Anirudh and Rizopoulos, Dimitris},
  journal={Annual Review of Statistics and Its Application},
  volume={6},
  number={1},
  pages={223--240},
  year={2019}
}

@article{liu2020semi,
  title={A semi-parametric joint latent class model with longitudinal and survival data},
  author={Liu, Yue and Lin, Ye and Zhou, Jianhui and Liu, Lei},
  journal={Statistics and Its Interface},
  volume={13},
  number={3},
  pages={411--422},
  year={2020}
}

@article{wong2022semiparametric,
  title={Semiparametric latent-class models for multivariate longitudinal and survival data},
  author={Wong, Kin Yau and Zeng, Donglin and Lin, DY},
  journal={Annals of Statistics},
  volume={50},
  number={1},
  pages={487},
  year={2022}
}

@article{barrett2019estimating,
  title={Estimating the association between blood pressure variability and cardiovascular disease: {A}n application using the {ARIC} Study},
  author={Barrett, Jessica K and Huille, Raphael and Parker, Richard and Yano, Yuichiro and Griswold, Michael},
  journal={Statistics in Medicine},
  volume={38},
  number={10},
  pages={1855--1868},
  year={2019}
}

@article{su2021risk,
  title={Risk factor identification in cystic fibrosis by flexible hierarchical joint models},
  author={Su, Weiji and Wang, Xia and Szczesniak, Rhonda D},
  journal={Statistical Methods in Medical Research},
  volume={30},
  number={1},
  pages={244--260},
  year={2021}
}

@article{hoffman2014no,
  title={The {N}o-{U}-{T}urn sampler: {A}daptively setting path lengths in {H}amiltonian {M}onte {C}arlo},
  author={Hoffman, Matthew D and Gelman, Andrew and others},
  journal={Journal of Machine Learning Research},
  volume={15},
  number={1},
  pages={1593--1623},
  year={2014}
}

@article{letenneur1994incidence,
  title={Incidence of dementia and {A}lzheimer's disease in elderly community residents of south-western {F}rance},
  author={Letenneur, Luc and Commenges, Daniel and Dartigues, Jean-Fran{\c{c}}ois and Barberger-Gateau, Pascale},
  journal={International Journal of Epidemiology},
  volume={23},
  number={6},
  pages={1256--1261},
  year={1994}
}

@Manual{proust2023package,
    title = {lcmm: Extended Mixed Models Using Latent Classes and Latent Processes},
    author = {Cecile Proust-Lima and Viviane Philipps and Amadou Diakite and Benoit Liquet},
    year = {2023},
    note = {R package version: 2.1.0},
    url = {https://cran.r-project.org/package=lcmm},
  }

@article{nguyen2023flash,
  title={{FLASH}: {A} fast joint model for longitudinal and survival data in high dimension},
  author={Nguyen, Van Tuan and Fermanian, Adeline and Guilloux, Agathe and Barbieri, Antoine and Zohar, Sarah and Jannot, Anne-Sophie and Bussy, Simon},
  journal={arXiv preprint arXiv:2309.03714},
  year={2023}
}

@article{alvares2023,
  title={A two-stage approach for {B}ayesian joint models: {R}educing complexity while maintaining accuracy},
  author={Alvares, Danilo and Leiva-Yamaguchi, Valeria},
  journal={Statistics and Computing},
  volume={33},
  number={5},
  pages={1--11},
  year={2023}
}

@article{rustand2024fast,
  title={Fast and flexible inference for joint models of multivariate longitudinal and survival data using integrated nested {L}aplace approximations},
  author={Rustand, Denis and Van Niekerk, Janet and Krainski, Elias Teixeira and Rue, H{\aa}vard and Proust-Lima, C{\'e}cile},
  journal={Biostatistics},
  volume={25},
  number={2},
  pages={429--448},
  year={2024}
}

@article{alvares2024,
  title={Bayesian survival analysis with {INLA}},
  author={Alvares, Danilo and Van Niekerk, Janet and Krainski, Elias Teixeira and Rue, H{\aa}vard and Rustand, Denis},
  journal={Statistics in Medicine},
  volume={43},
  number={20},
  pages={3975--4010},
  year={2024}
}

@article{gomez2018markov,
  title={Markov chain {M}onte {C}arlo with the integrated nested {L}aplace approximation},
  author={G{\'o}mez-Rubio, Virgilio and Rue, H{\aa}vard},
  journal={Statistics and Computing},
  volume={28},
  pages={1033--1051},
  year={2018}
}

@article{sun2024penalized,
  title={Penalized joint models of high-dimensional longitudinal biomarkers and a survival outcome},
  author={Sun, Jiehuan and Basu, Sanjib},
  journal={The Annals of Applied Statistics},
  volume={18},
  number={2},
  pages={1490--1505},
  year={2024}
}

@article{sivula2020uncertainty,
  title={Uncertainty in {B}ayesian leave-one-out cross-validation based model comparison},
  author={Sivula, Tuomas and Magnusson, M{\aa}ns and Matamoros, Asael Alonzo and Vehtari, Aki},
  journal={arXiv preprint arXiv:2008.10296},
  year={2020}
}

@article{gorus2008reaction,
  title={Reaction times and performance variability in normal aging, mild cognitive impairment, and {A}lzheimer's disease},
  author={Gorus, Ellen and De Raedt, Rudi and Lambert, Margareta and Lemper, Jean-Claude and Mets, Tony},
  journal={Journal of Geriatric Psychiatry and Neurology},
  volume={21},
  number={3},
  pages={204--218},
  year={2008}
}

@article{hultsch2000intraindividual,
  title={Intraindividual variability in cognitive performance in older adults: {C}omparison of adults with mild dementia, adults with arthritis, and healthy adults},
  author={Hultsch, David F and MacDonald, Stuart WS and Hunter, Michael A and Levy-Bencheton, Judi and Strauss, Esther},
  journal={Neuropsychology},
  volume={14},
  number={4},
  pages={588},
  year={2000}
}

\pagebreak

\captionsetup[figure]{name=Supplementary Figure}
\captionsetup[table]{name=Supplementary Table}
\setcounter{figure}{0}
\setcounter{table}{0}

\begin{center}
    {\LARGE Supplementary material to the paper \\ \textbf{ Bayesian shared parameter joint models for heterogeneous populations}} \bigskip \\
    \normalsize \textbf{Sida Chen}, \textbf{Danilo Alvares}, \textbf{Marco Palma}, \textbf{Jessica K. Barrett} \\
    \normalsize MRC Biostatistics Unit, University of Cambridge, U.K.
\end{center}

\appendix

\begin{figure}[H]
    \centering
    \includegraphics[width=0.8\textwidth]{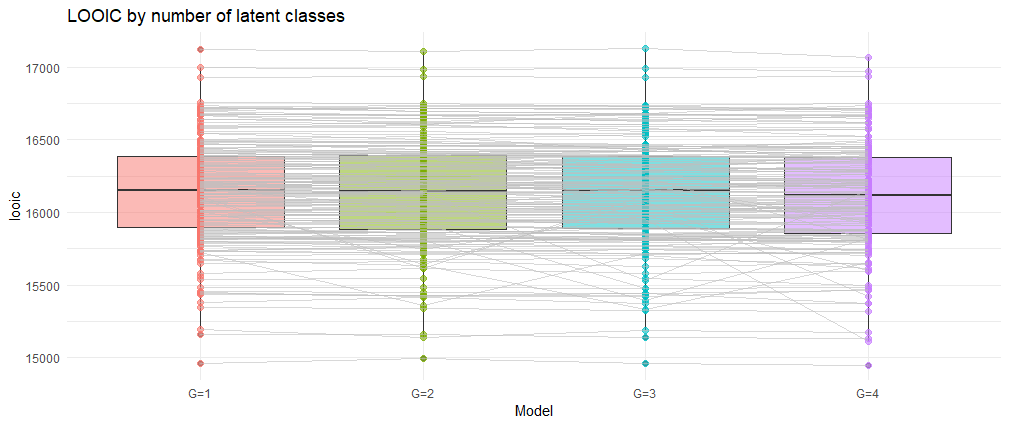}
    \includegraphics[width=0.8\textwidth]{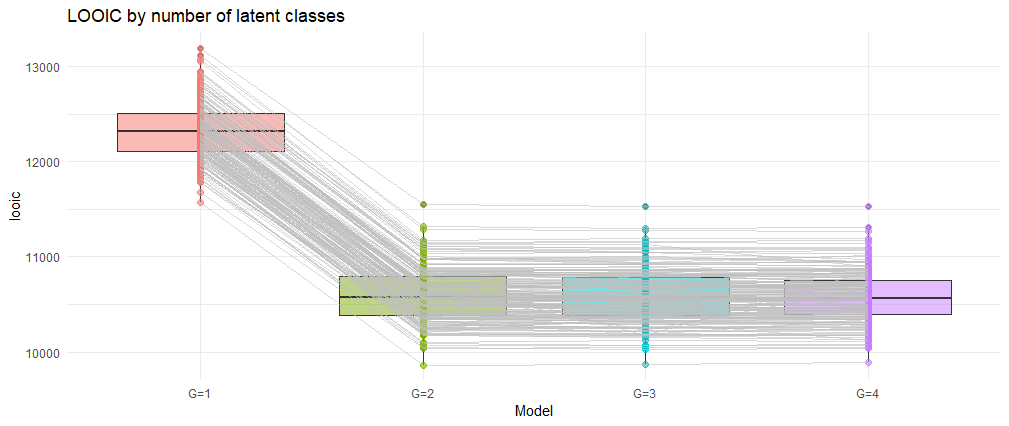}
    \includegraphics[width=0.8\textwidth]{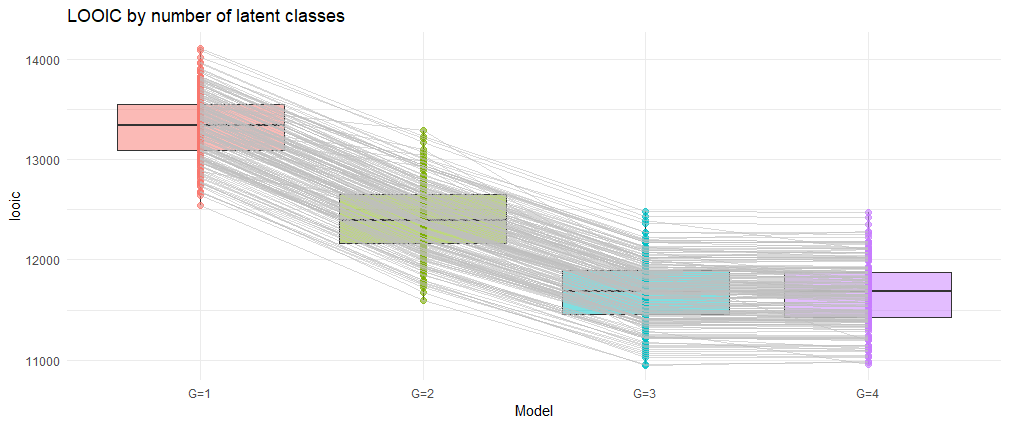}
    \caption{Boxplot summary of LOOIC (on the deviance scale) computed for candidate values of $G$ based on 200 data replications for Scenario 1 (top panel), Scenario 2 (middle panel), and Scenario 3 (bottom panel) from simulation Setting I. Points representing LOOIC for specific data replications are superimposed on the boxplots and are connected by lines when based on the same data replication. }
    \label{fig:sim_looicplots}
\end{figure}

\begin{figure}[H]
    \centering
    \includegraphics[width=0.8\textwidth]{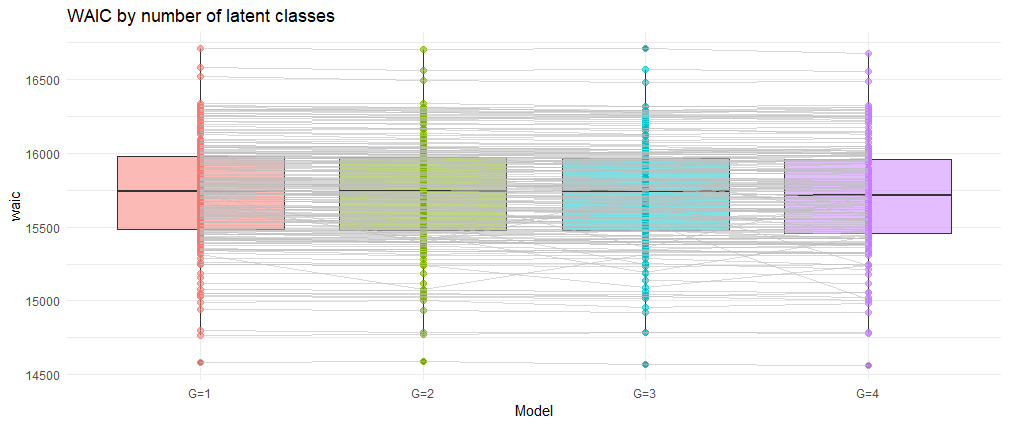}
    \includegraphics[width=0.8\textwidth]{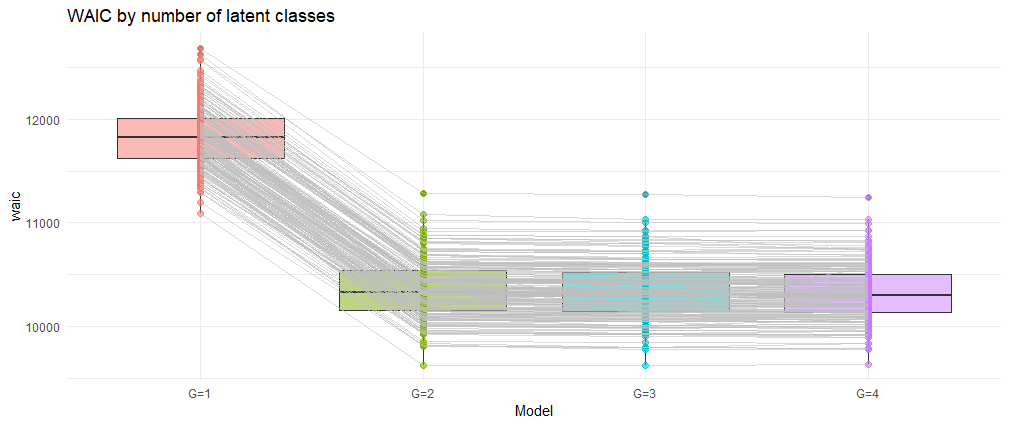}
    \includegraphics[width=0.8\textwidth]{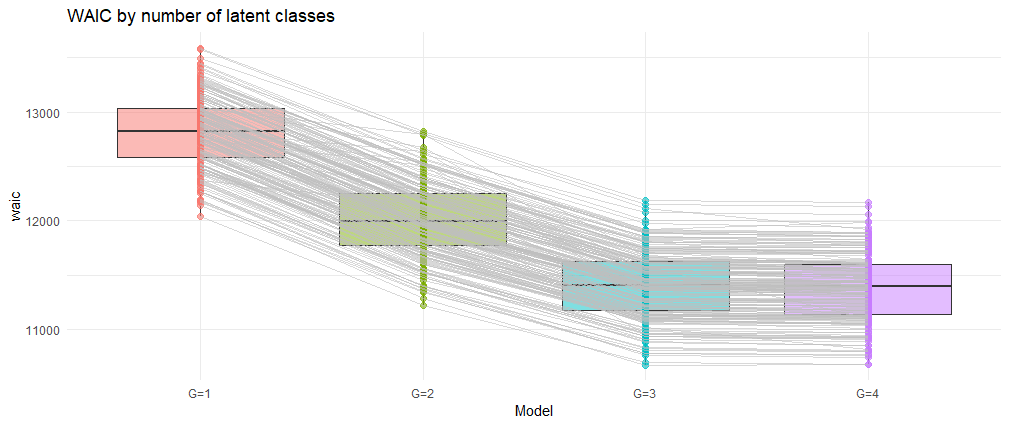}
    \caption{Boxplot summary of WAIC (on the deviance scale) computed for candidate values of $G$ based on 200 data replications for Scenario 1 (top panel), Scenario 2 (middle panel), and Scenario 3 (bottom panel) from simulation Setting I. Points representing WAIC for specific data replications are superimposed on the boxplots and are connected by lines when based on the same data replication.}
    \label{fig:sim_waicplots}
\end{figure}

\begin{table}[ht]
\centering
\begin{tabular}{ccccc}
  \toprule
  \textbf{Parameter} & \textbf{True Value} & \textbf{Bias} & \textbf{SD} & \textbf{Coverage (\%)} \\ 
  \midrule
  $\beta_{1,0}$ & 8.03 & 0.009 & 0.143 & 95.5 \\ 
  $\beta_{1,1}$ & -0.16 & 0.000 & 0.023 & 95.5 \\ 
  $\beta_{1,2}$ & -5.86 & -0.009 & 0.207 & 96.5 \\ 
  $\beta_{2,0}$ & -8.03 & -0.004 & 0.051 & 93.0 \\ 
  $\beta_{2,1}$ & 0.46 & 0.009 & 0.094 & 94.0 \\ 
  $\beta_{2,2}$ & 12.2 & -0.003 & 0.072 & 93.0 \\ 
  $\beta_{3,0}$ & 0.03 & 0.001 & 0.046 & 96.5 \\ 
  $\beta_{3,1}$ & -0.01 & -0.002 & 0.035 & 93.5 \\ 
  $\beta_{3,2}$ & -1.96 & 0.001 & 0.061 & 97.0 \\ 
  $\gamma_{1,0}$ & -4.85 & -0.165 & 0.807 & 93.5 \\ 
  $\gamma_{1,1}$ & -0.02 & -0.002 & 0.011 & 95.0 \\ 
  $\gamma_{2,0}$ & -4.85 & -0.079 & 0.300 & 96.0 \\ 
  $\gamma_{2,1}$ & 0.09 & 0.001 & 0.006 & 96.5 \\ 
  $\gamma_{3,0}$ & 2.85 & 0.036 & 0.228 & 94.5 \\ 
  $\gamma_{3,1}$ & -0.12 & -0.002 & 0.007 & 93.5 \\ 
  $\alpha_{1}$ & 0.38 & 0.014 & 0.066 & 93.5 \\ 
  $\alpha_{2}$ & 0.08 & 0.002 & 0.011 & 92.0 \\ 
  $\alpha_{3}$ & 0.58 & 0.010 & 0.034 & 96.5 \\ 
  $\xi_{1}$ & 1.8 & 0.065 & 0.228 & 95.0 \\ 
  $\xi_{2}$ & 1.4 & 0.020 & 0.066 & 95.5 \\ 
  $\xi_{3}$ & 1.8 & 0.032 & 0.088 & 91.5 \\ 
  $\sigma^{2}_{1}$ & 0.4761 & 0.002 & 0.030 & 95.0 \\ 
  $\sigma^{2}_{2}$ & 0.4761 & -0.006 & 0.029 & 94.0 \\ 
  $\sigma^{2}_{3}$ & 0.4761 & -0.001 & 0.015 & 95.5 \\ 
  $\Sigma_{1,11}$ & 0.87 & 0.025 & 0.166 & 93.5 \\ 
  $\Sigma_{1,22}$ & 0.02 & 0.003 & 0.007 & 94.0 \\ 
  $\Sigma_{2,11}$ & 0.02 & 0.021 & 0.020 & 92.5 \\ 
  $\Sigma_{2,22}$ & 0.91 & 0.013 & 0.118 & 97.0 \\ 
  $\Sigma_{3,11}$ & 0.28 & 0.006 & 0.035 & 94.0 \\ 
  $\Sigma_{3,22}$ & 0.31 & 0.002 & 0.028 & 95.0 \\ 
  \bottomrule
\end{tabular}
\caption{Results for simulation Setting I (Scenario 3) with G fixed at 3. For each parameter, bias and standard deviation (SD) are evaluated based on the posterior mean obtained from 200 data replications, and coverage is the proportion of times the 95\% credible interval contains the true parameter value across the 200 replications.}
\label{tab:sim_m1s3}
\end{table}

\begin{table}[h!]
\centering
\begin{tabular}{cccccccc}
\toprule
\multirow{2}{*}{\textbf{Parameter}} & \multirow{2}{*}{\textbf{True Value}} & \multicolumn{3}{c}{\textbf{Model 1}} & \multicolumn{3}{c}{\textbf{Model 2}} \\
\cmidrule(lr){3-5} \cmidrule(lr){6-8}
& & Bias & SD & Coverage (\%) & Bias & SD & Coverage (\%) \\
\midrule
$\beta_{1,0}$ & 8.03   & -0.006 & 0.085 & 93.5  & -0.006 & 0.084 & 94.0 \\
$\beta_{1,1}$ & -0.16  &  0.000 & 0.014 & 94.0  &  0.000 & 0.014 & 95.0 \\
$\beta_{1,2}$ & -5.86  &  0.018 & 0.127 & 94.5  &  0.019 & 0.124 & 96.0 \\
$\beta_{2,0}$ & -8.03  &  0.001 & 0.032 & 95.0  &  0.001 & 0.033 & 94.5 \\
$\beta_{2,1}$ & 0.46   & -0.006 & 0.063 & 93.5  & -0.006 & 0.063 & 94.5 \\
$\beta_{2,2}$ & 12.2   & -0.005 & 0.048 & 93.5  & -0.005 & 0.049 & 93.0 \\
$\gamma_{1,0}$  & -4.85  & -0.024 & 0.428 & 94.5  & -0.023 & 0.433 & 93.5 \\
$\gamma_{1,1}$  & -0.02  & -0.001 & 0.005 & 95.5  & -0.001 & 0.005 & 96.0 \\
$\gamma_{2,0}$  & -4.85  & -0.012 & 0.201 & 97.5  & -0.010 & 0.201 & 96.5 \\
$\gamma_{2,1}$  & 0.09   &  0.000 & 0.004 & 98.0  &  0.000 & 0.004 & 97.0 \\
$\alpha_{1}$   & 0.38   &  0.004 & 0.037 & 94.0  &  0.004 & 0.036 & 93.0 \\
$\alpha_{2}$   & 0.08   &  0.000 & 0.007 & 95.5  &  0.000 & 0.007 & 95.5 \\
$\xi_{1}$  & 1.8    &  0.029 & 0.124 & 92.5  &  0.028 & 0.124 & 92.5 \\
$\xi_{2}$  & 1.4    &  0.005 & 0.046 & 95.5  &  0.005 & 0.047 & 94.0 \\
$\sigma^{2}_{1}$ & 0.4761 & -0.001 & 0.019 & 94.0  & -0.001 & 0.019 & 95.0 \\
$\sigma^{2}_{2}$ & 0.4761 & -0.009 & 0.021 & 91.0  & -0.008 & 0.021 & 91.5 \\
$\Sigma_{1,11}$      & 0.87   &  0.006 & 0.095 & 92.5  &  0.006 & 0.095 & 94.0 \\
$\Sigma_{1,22}$      & 0.02   &  0.001 & 0.004 & 92.0  &  0.001 & 0.005 & 93.0 \\
$\Sigma_{2,11}$      & 0.02   &  0.013 & 0.017 & 91.0  &  0.012 & 0.017 & 91.0 \\
$\Sigma_{2,22}$      & 0.91   &  0.006 & 0.083 & 98.0  &  0.007 & 0.084 & 97.5 \\
\multicolumn{2}{c}{} & \multicolumn{3}{c}{Classification accuracy} & \multicolumn{3}{c}{Classification accuracy} \\
\multicolumn{2}{c}{} & \multicolumn{3}{c}{ 98.8 (98.6, 99.0)} & \multicolumn{3}{c}{98.8 (98.4, 99.0)} \\
\bottomrule
\end{tabular}
\caption{Results for simulation Setting II (Scenario 1). For each parameter, bias and standard deviation (SD) are evaluated based on the posterior mean obtained from 200 data replications, and coverage is the proportion of times the 95\% credible interval contains the true parameter value across the 200 replications. Classification accuracy is summarized using the median and interquartile range (shown in brackets), with accuracy for each replication calculated as the proportion of correctly classified subjects.
Model 1: model with homogeneous mixture weights (ignoring covariate effects on the class membership). Model 2: model with covariate dependent mixture weights (use gender and age only).}
\label{tab:sim_m2s1}
\end{table}

\begin{table}[h!]
\centering
\begin{tabular}{cccccccc}
\toprule
\multirow{2}{*}{\textbf{Parameter}} & \multirow{2}{*}{\textbf{True Value}} & \multicolumn{3}{c}{\textbf{Model 1}} & \multicolumn{3}{c}{\textbf{Model 2}} \\
\cmidrule(lr){3-5} \cmidrule(lr){6-8}
& & Bias & SD & Coverage (\%) & Bias & SD & Coverage (\%) \\
\midrule
$\beta_{1,0}$    & 8.03   & -0.005 & 0.085 & 93.0  & -0.006 & 0.085 & 92.0 \\
$\beta_{1,1}$    & -0.16  &  0.001 & 0.012 & 95.0  &  0.001 & 0.012 & 96.0 \\
$\beta_{1,2}$    & -5.86  &  0.004 & 0.121 & 92.5  &  0.005 & 0.121 & 94.0 \\
$\beta_{2,0}$    & -8.03  &  0.005 & 0.038 & 94.0  &  0.005 & 0.038 & 94.0 \\
$\beta_{2,1}$    & 0.46   &  0.000 & 0.069 & 92.0  &  0.000 & 0.070 & 93.5 \\
$\beta_{2,2}$    & 12.2   & -0.007 & 0.053 & 96.0  & -0.008 & 0.053 & 95.0 \\
$\gamma_{1,0}$     & -4.85  & -0.078 & 0.376 & 94.5  & -0.076 & 0.377 & 95.0 \\
$\gamma_{1,1}$     & -0.02  &  0.000 & 0.005 & 95.5  &  0.000 & 0.005 & 95.5 \\
$\gamma_{2,0}$     & -4.85  & -0.036 & 0.243 & 95.0  & -0.036 & 0.245 & 94.5 \\
$\gamma_{2,1}$     & 0.09   &  0.001 & 0.005 & 95.0  &  0.001 & 0.005 & 94.0 \\
$\alpha_{1}$      & 0.38   &  0.005 & 0.031 & 95.5  &  0.004 & 0.031 & 95.5 \\
$\alpha_{2}$      & 0.08   &  0.001 & 0.008 & 95.0  &  0.001 & 0.007 & 96.0 \\
$\xi_{1}$     & 1.8    &  0.030 & 0.111 & 93.5  &  0.029 & 0.112 & 95.5 \\
$\xi_{2}$     & 1.4    &  0.010 & 0.051 & 94.0  &  0.010 & 0.051 & 94.0 \\
$\sigma^{2}_{1}$ & 0.4761 &  0.000 & 0.015 & 95.5  &  0.001 & 0.015 & 97.0 \\
$\sigma^{2}_{2}$ & 0.4761 & -0.005 & 0.022 & 96.0  & -0.005 & 0.022 & 95.5 \\
$\Sigma_{1,11}$         & 0.87   &  0.008 & 0.072 & 96.0  &  0.008 & 0.072 & 97.0 \\
$\Sigma_{1,22}$         & 0.02   &  0.001 & 0.003 & 93.5  &  0.001 & 0.003 & 94.0 \\
$\Sigma_{2,11}$         & 0.02   &  0.014 & 0.018 & 87.5  &  0.015 & 0.018 & 91.5 \\
$\Sigma_{2,22}$         & 0.91   &  0.017 & 0.085 & 97.0  &  0.015 & 0.085 & 96.0 \\
\multicolumn{2}{c}{} & \multicolumn{3}{c}{Classification accuracy} & \multicolumn{3}{c}{Classification accuracy} \\
\multicolumn{2}{c}{} & \multicolumn{3}{c}{98.7 (98.4, 98.9)} & \multicolumn{3}{c}{98.7 (98.4, 98.9)} \\
\bottomrule
\end{tabular}
\caption{Results for simulation Setting II (Scenario 2). The settings are the same as those in Table \ref{tab:sim_m2s1}.}
\label{tab:sim_m2s2}
\end{table}

\end{document}